\newif\iftechreport
\techreporttrue %

\iftechreport
\documentclass[acmsmall,nonacm]{acmart}
\else
\documentclass[acmsmall]{acmart}
\fi

\AtBeginDocument{%
  }

\setcopyright{acmlicensed}
\copyrightyear{2018}
\acmYear{2018}
\acmDOI{XXXXXXX.XXXXXXX}

\acmConference[Conference acronym 'XX]{Make sure to enter the correct
  conference title from your rights confirmation email}{June 03--05,
  2018}{Woodstock, NY}
\acmISBN{978-1-4503-XXXX-X/18/06}

\usepackage{xcolor}
\usepackage[a-2b]{pdfx}
\usepackage{hyperref}
\usepackage{subcaption}
\usepackage{xspace}
\usepackage{amsmath}
\usepackage[compress]{cleveref}
\usepackage{array, caption, booktabs}
\usepackage[linesnumbered,ruled,noend]{algorithm2e}
\usepackage{setspace}
\usepackage[title]{appendix}
\usepackage{bbm}
\usepackage{makecell}
\usepackage{multirow}
\usepackage{subcaption}
\usepackage{listings}
\usepackage{color}
\usepackage{enumitem}
\usepackage{graphicx}
\usepackage{balance}
\usepackage[normalem]{ulem}
\usepackage{hhline}
\usepackage{lipsum}
\usepackage[frozencache,cachedir=.]{minted}
\usepackage[most]{tcolorbox}
\usepackage{ifthen}

\definecolor{kwgreen}{rgb}{0.03,0.5,0}
\definecolor{dkgreen}{rgb}{0,0.6,0}
\definecolor{gray}{rgb}{0.5,0.5,0.5}
\definecolor{mauve}{rgb}{0.58,0,0.82}
\crefname{section}{Section}{Sections}
\Crefname{section}{Section}{Sections}
\crefname{figure}{Figure}{Figures}
\Crefname{figure}{Figure}{Figures}
\crefname{subfigure}{Figure}{Figures}
\Crefname{subfigure}{Figure}{Figures}
\crefname{lstlisting}{Listing}{Listings}
\Crefname{lstlisting}{Listing}{Listings}
\crefrangelabelformat{subfigure}{#3#1#4--#5(\crefstripprefix{#1}{#2}#6}

\definecolor{hlc1}{cmyk}{0.15,0.01,0.01,0.0}
\definecolor{hlc2}{cmyk}{0.12,0.15,0.0,0.0}
\definecolor{hlc3}{cmyk}{0.10,0.01,0.29,0.0}
\definecolor{hlc4}{cmyk}{0.02, 0.10, 0.20, 0.0}
\definecolor{hlc5}{cmyk}{0.20, 0.0, 0.15, 0.0}
\definecolor{hlc6}{cmyk}{0.0, 0.20, 0.25, 0.0}
\definecolor{hlc7}{cmyk}{0.20, 0.07, 0.0, 0.0}
\definecolor{hlc8}{cmyk}{0.0, 0.32, 0.25, 0.0}
\definecolor{hlc9}{cmyk}{0.25, 0.20, 0.0, 0.0}

\newcommand*\highlightht{2.8ex}
\newcommand*\highlightdp{-.8ex}
\newcommand*\highlightwd{0.2ex}
\newcommand{\highlight}[2]{\highlightcommon{#1}{#2}}
\newcommand{\highlightcommon}[1]
  {%
    \bgroup
    \markoverwith
      {\textcolor{#1}{\smash{\rule[\highlightdp]{\highlightwd}{\highlightht}}}}%
    \ULon
  }

 \newcommand{\revision}[1]{{#1}}

\newcommand{\system}{VOCAL-UDF\xspace}

\begin{document}
\iftechreport
\title{Self-Enhancing Video Data Management System for Compositional Events with Large Language Models [Technical Report]}
\renewcommand{\shorttitle}{Self-Enhancing Video Data Management System for Compositional Events with LLMs [Technical Report]}
\fi 

\author{Enhao Zhang}
\affiliation{%
  \institution{University of Washington}
  \city{Seattle}
  \country{USA}
}
\email{enhaoz@cs.washington.edu}

\author{Nicole Sullivan}
\affiliation{%
  \institution{University of Washington}
  \city{Seattle}
  \country{USA}
}
\email{nsulliv@cs.washington.edu}

\author{Brandon Haynes}
\affiliation{%
  \institution{Microsoft Gray Systems Lab}
  \city{Redmond}
  \country{USA}
}
\email{brandon.haynes@microsoft.com}

\author{Ranjay Krishna}
\affiliation{%
  \institution{University of Washington}
  \city{Seattle}
  \country{USA}
}
\email{ranjay@cs.washington.edu}

\author{Magdalena Balazinska}
\affiliation{%
  \institution{University of Washington}
  \city{Seattle}
  \country{USA}
}
\email{magda@cs.washington.edu}

\begin{abstract}
Complex video queries can be answered by decomposing them into modular subtasks. 
However, existing video data management systems assume the existence of predefined modules for each subtask.
\iftechreport
We introduce \system\footnote{The source code is available at \url{https://github.com/uwdb/VOCAL-UDF}},
\fi
a novel self-enhancing system that supports compositional queries over videos without the need for predefined modules. 
\system automatically identifies and constructs missing modules and encapsulates them as user-defined functions (UDFs), thus expanding its querying capabilities. 
To achieve this, we formulate a unified UDF model that leverages large language models (LLMs) to aid in new UDF generation. 
\system handles a wide range of concepts by supporting both \textit{program-based UDFs} (i.e., Python functions generated by LLMs) and \textit{distilled-model UDFs} (lightweight vision models distilled from strong pretrained models). To resolve the inherent ambiguity in user intent, \system generates multiple candidate UDFs and uses active learning to efficiently select the best one. 
With the self-enhancing capability, \system significantly improves query performance across three video datasets.
\end{abstract}

\maketitle

\begin{sloppypar}
    
\newcommand{\teaserFigure}{
    \begin{figure}[t!]
        \centering
        \includegraphics[width=0.9\columnwidth]{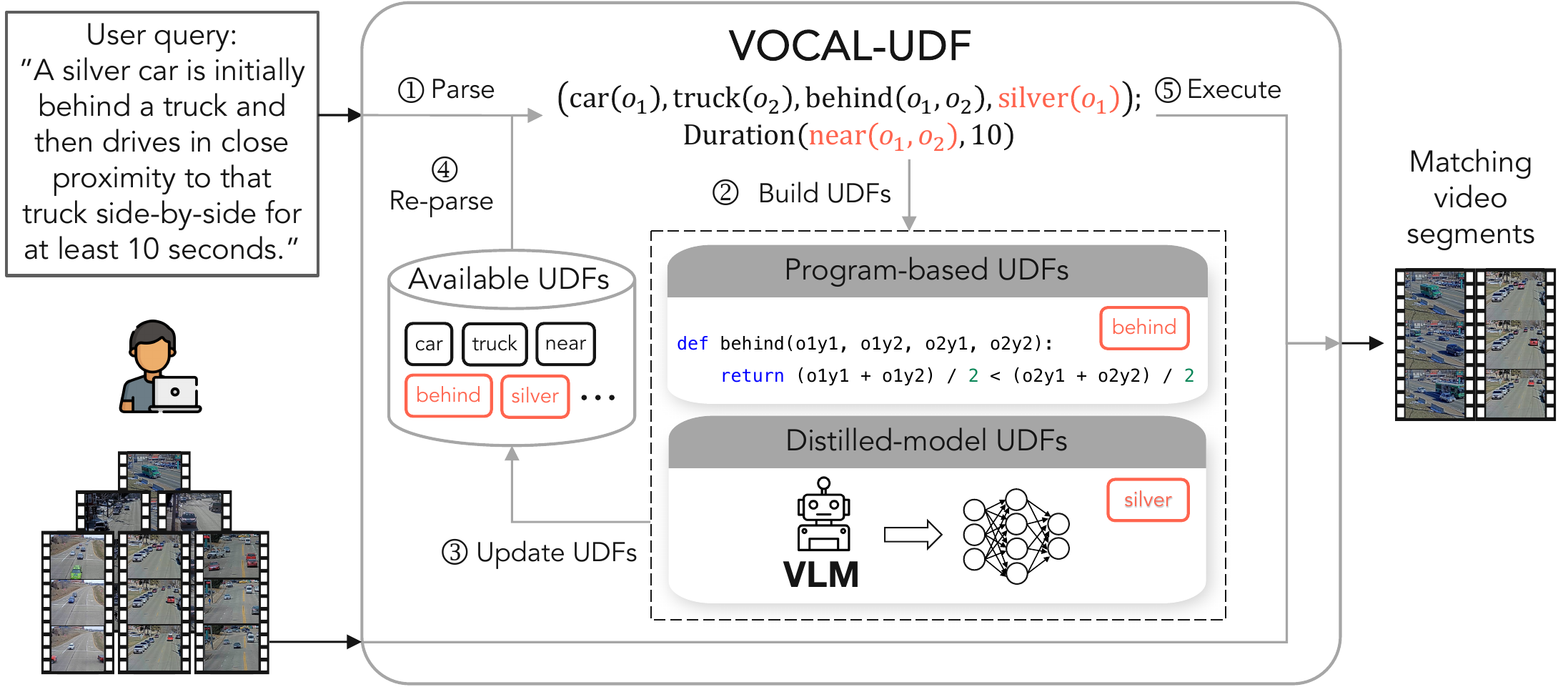}
        \caption{Given a video dataset and a user query in natural language, \system \textcircled{\small 1} parses the query into a DSL notation. If the query contains predicates that existing UDFs cannot answer, \system \textcircled{\small 2} automatically builds new UDFs, \textcircled{\small 3} updates its available UDF list, \textcircled{\small 4} reparses the query, and \textcircled{\small 5} executes the query to return matching video segments. \system supports both program-based UDFs (i.e., Python functions) and distilled-model UDFs (i.e., ML models) for diverse concepts.}
        \label{fig:teaser}
    \end{figure}
}

\newcommand{\systemFigure}{
    \begin{figure}[t!]
        \centering
        \includegraphics[width=0.5\columnwidth]{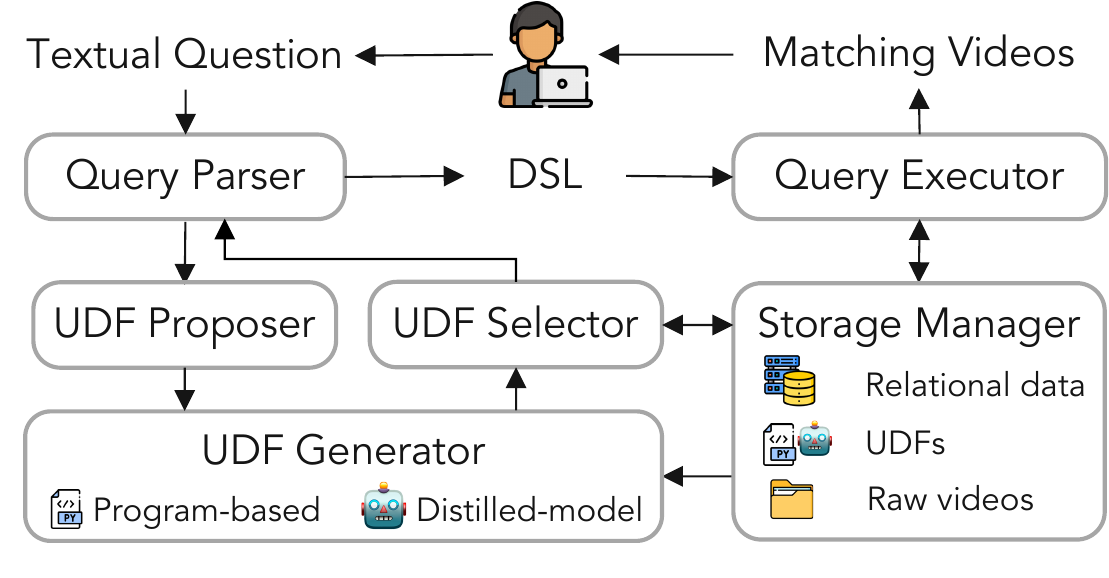}
        \caption{\system system overview.}
        \label{fig:system}
    \end{figure}
}

\newcommand{\programCandidatesFigure}{
    \begin{figure}[t!]
        \centering
        \includegraphics[width=0.7\columnwidth]{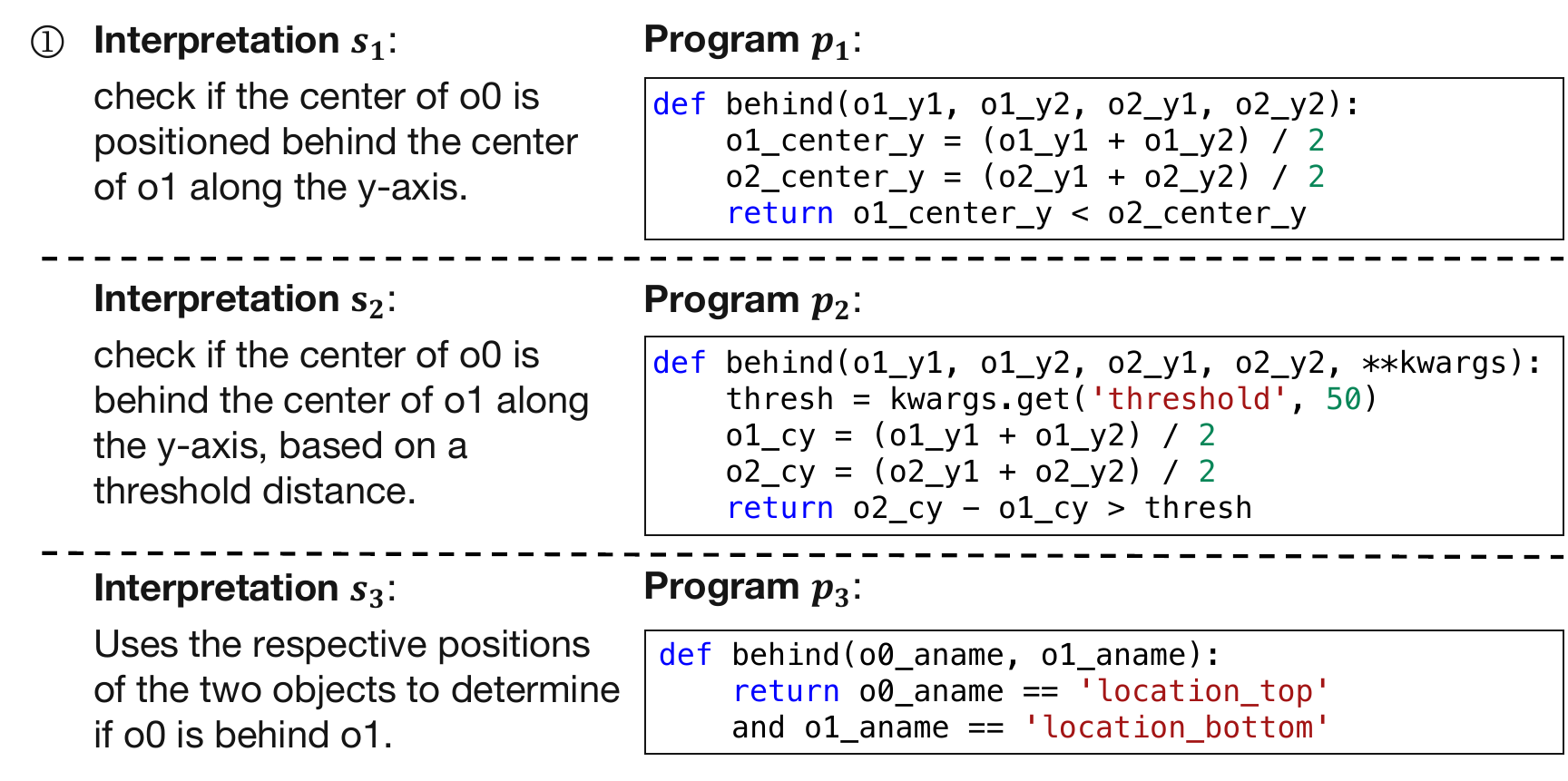}
        \caption{Program candidates, using \texttt{behind} as an example.}
        \label{fig:program_candidates}
    \end{figure}
}

\newcommand{\dataLabelingFigure}{
    \begin{figure}[t!]
        \centering
        \includegraphics[width=0.7\columnwidth]{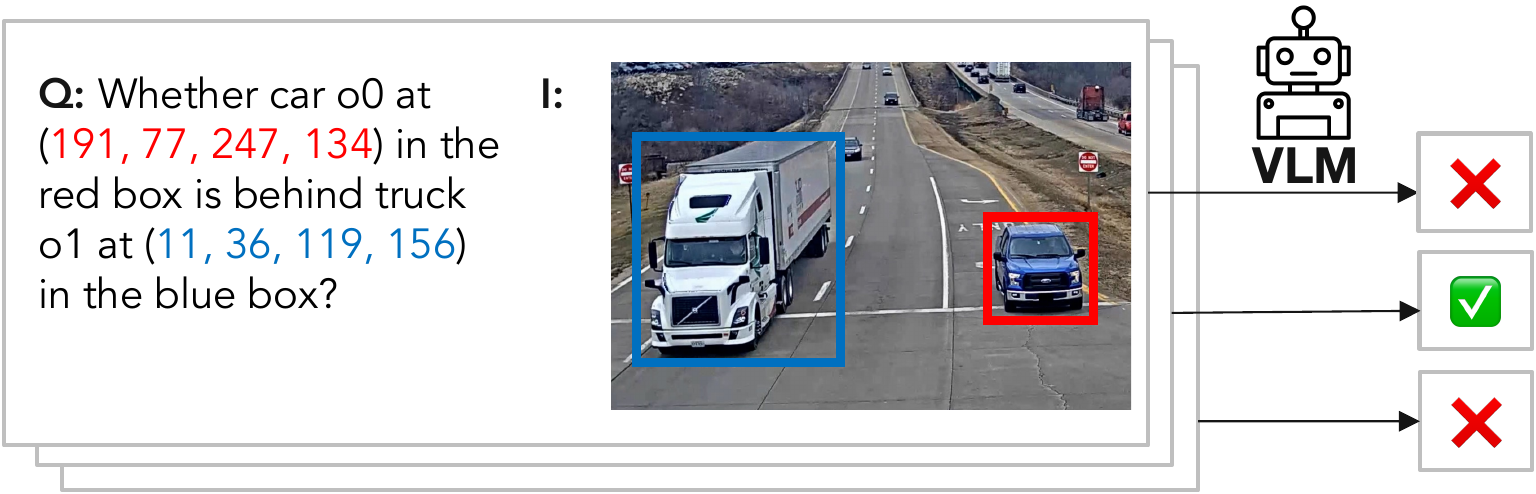}
        \caption{Data labeling by a VLM, using \texttt{behind} as an example.}
        \label{fig:data_labeling}
    \end{figure}
}

\newcommand{\evalAllFigure}{
    \begin{figure}[t!]
        \centering
        \includegraphics[height=0.8cm]{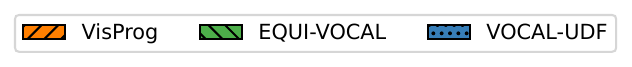}
        \includegraphics[width=0.8\columnwidth]{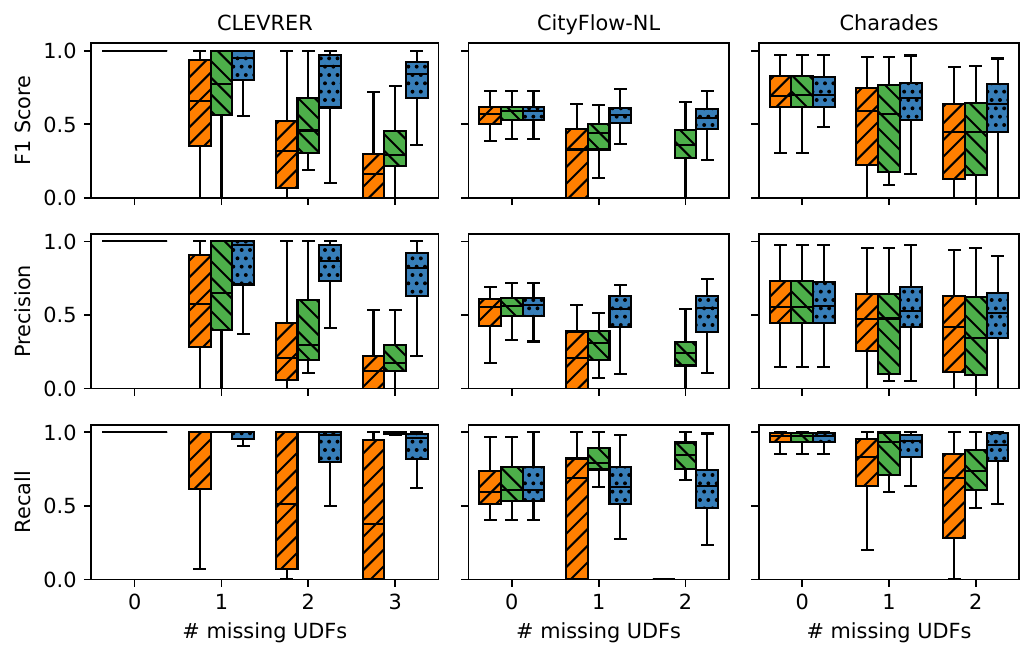}
        \caption{\revision{End-to-end performance (F1 score, precision, recall) of generated queries with various number of missing UDFs.}}
        \label{fig:eval_all}
    \end{figure}
}

\newcommand{\vanillaLlmFigure}{
    \begin{figure}[t!]
        \begin{subfigure}{0.48\columnwidth}
        \centering
            \includegraphics[height=12em]{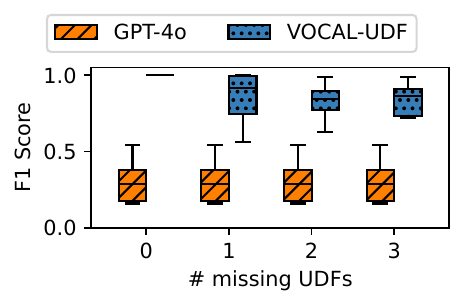}
            \caption{Vanilla LLM vs. \system.}
            \label{subfig:llm_baseline}
        \end{subfigure}
        \begin{subfigure}{0.48\columnwidth}
        \centering
            \includegraphics[height=12em]{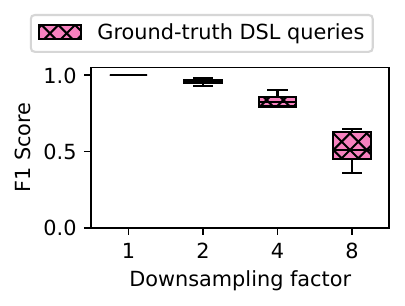}
            \caption{\revision{Downsampling reduces F1 scores.}}
            \label{subfig:sampling_rate}
        \end{subfigure}
        \caption{\revision{(a) F1 scores of the vanilla LLM approach and \system on CLEVRER under a simplified setting. (b) F1 scores of running ground-truth DSL queries over sampled videos.}}
        \label{fig:llm_baseline}
    \end{figure}
}

\newcommand{\EfficiencyFigure}{
    \begin{figure}[t!]
        \centering
        \begin{subfigure}{\columnwidth}
            \centering
            \includegraphics[width=0.9\columnwidth]{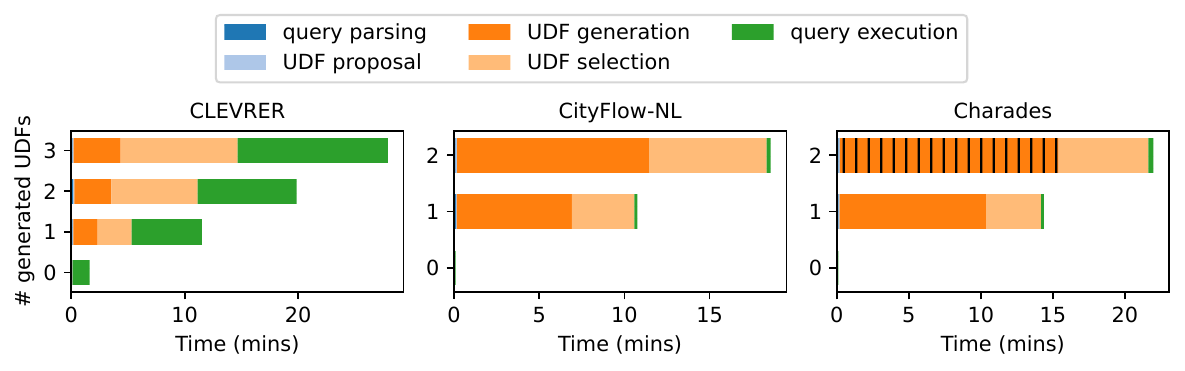}
            \caption{Total time.}
            \label{subfig:total_time}
        \end{subfigure}
        \begin{subfigure}{\columnwidth}
            \centering
            \includegraphics[width=0.9\columnwidth]{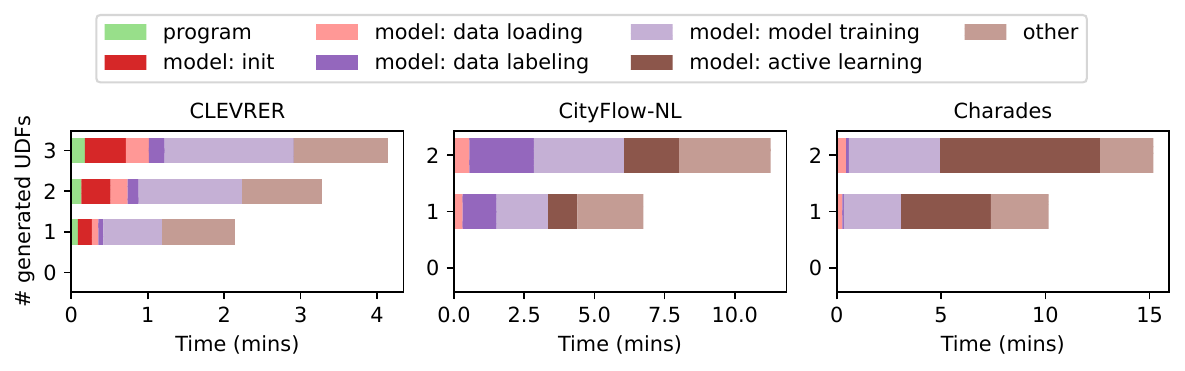}
            \caption{UDF generation time.}
            \label{subfig:udf_generation_time}
        \end{subfigure}
        \begin{subfigure}{\columnwidth}
            \centering
            \includegraphics[width=0.9\columnwidth]{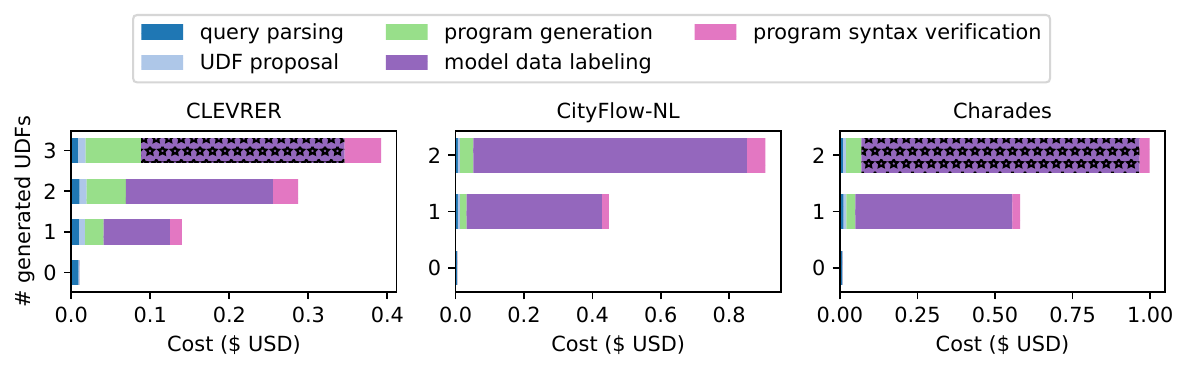}
            \caption{Monetary cost of OpenAI API invocation.}
            \label{subfig:cost}
        \end{subfigure}
        \caption{\revision{Detailed breakdown of efficiency and cost.}}
        \label{fig:system_efficiency}
    \end{figure}
}

\newcommand{\intentAmbiguityFigure}{
    \begin{figure}[t!]
        \centering
        \includegraphics[width=0.7\columnwidth]{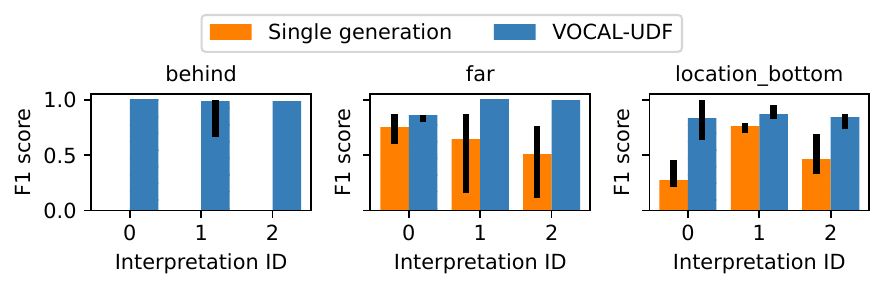}
        \caption{\revision{Median F1 scores of \system vs. a single-generation baseline on queries with one predicate under various interpretations, each run 20 times. Error bars show the IQR.}}
        \label{fig:intent_ambiguity}
    \end{figure}
}

\newcommand{\udfSelectionFigure}{
    \begin{figure}[t!]
        \centering
        \includegraphics[width=0.6\columnwidth]{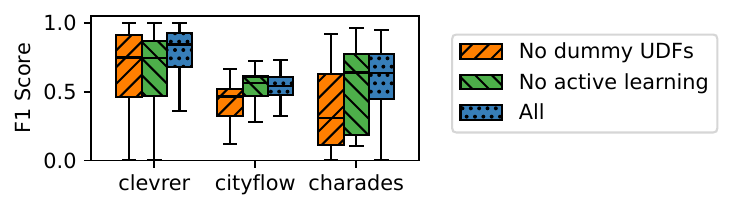}
        \caption{F1 scores of \system without active learning and without dummy UDFs in UDF selection.}
        \label{fig:udf_selection}
    \end{figure}
}

\newcommand{\initialLabelingFigure}{
    \begin{figure}[t!]
        \centering
        \includegraphics[width=0.7\columnwidth]{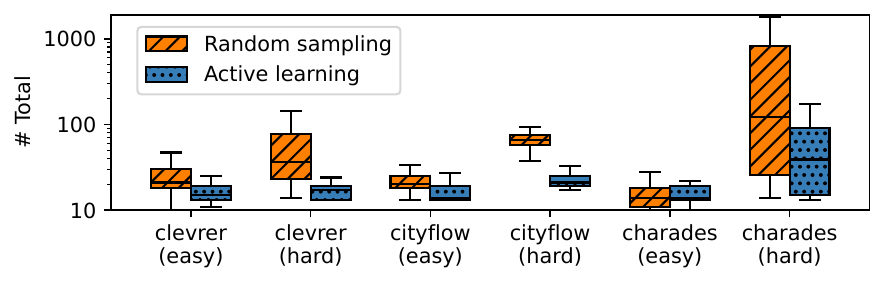}
        \caption{\revision{Number of samples required to get at least 10 positive examples (y-axis in log scale; lower is better).}}
        \label{fig:init_label}
    \end{figure}
}
\newcommand{\relationalSchema}{
    \begin{table}[t!]
        \centering
        \caption{Relational schema representation of data model.}
        \label{table:relational_schema}
        {\renewcommand{\arraystretch}{1}%
        \begin{tabular}{|l|}
          \hline
          \texttt{Frames}(vid, fid, pixels) \\ \hline
          \texttt{Objects}(vid, fid, oid, oname, $x_1$, $y_1$, $x_2$, $y_2$) \\ \hline
          \texttt{Relationships}(vid, fid, rid, $\textrm{oid}_1$, rname, $\textrm{oid}_2$) \\ \hline
          \texttt{Attributes}(vid, fid, oid, aname) \\ \hline
        \end{tabular}}
    \end{table}
}

\newcommand{\relationalView}{
    \begin{table}[t!]
        \centering
        \caption{Relational views for query execution with UDFs.}
        \label{table:relational_view}
        {\renewcommand{\arraystretch}{1}%
        \begin{tabular}{|l|}
          \hline
          \texttt{ObjView}(vid, video$\_$pixels) \\ \hline
          \texttt{RelView}(vid, fid, pixels, rid, o1$\_$o2$\_$rnames, o2$\_$o1$\_$rnames, \\
          o1$\_$oid, o1$\_$x1, o1$\_$y1, o1$\_$x2, o1$\_$y2, o1$\_$anames, \\
          o2$\_$oid, o2$\_$x1, o2$\_$x2, o2$\_$y1, o2$\_$y2, o2$\_$anames) \\ \hline
          \texttt{AttrView}(vid, fid, pixels, oid, oname, x1, y1, x2, y2, anames) \\ \hline
        \end{tabular}}
    \end{table}
}

\newcommand{\datasetTable}{
    \begin{table*}[t]
    \centering
    \caption{Dataset summary. $^\dagger$Charades contains 15 semantic relationships, but only 9 of them are observed in test queries. $^\ddagger$For attributes in \textsc{CLEVRER}, colors, shapes, and materials are extracted by Mask R-CNN, while other attributes are programs.}
    \footnotesize
    \setlength{\tabcolsep}{4pt}
    \begin{tabular}{c|cccc|ccc|p{0.14\textwidth}p{0.18\textwidth}|cccc}
        \hline
        \multirow{2}{*}{Dataset} & \multicolumn{4}{c|}{Scene graph annotations} & \multicolumn{3}{c|}{Predefined UDF types} & \multicolumn{2}{c|}{In-context examples} & \multicolumn{4}{c}{Query complexity} \\ \cline{2-14}
        & Obj & \makecell{Sp. \\ Rel} & \makecell{Sem. \\ Rel} & Attr & \makecell{Sp. \\ Rel} & \makecell{Sem. \\ Rel} & Attr & Base UDFs & Supplemental UDFs & \# Preds & \# Vars & \makecell{\# Reg. \\ graphs} & Duration  \\ \hline
        \textsc{CLEVRER} & 1 & 6 & --- & 17 & Prog. & --- & \makecell{Prog. \& \\ Obj. Det.}$^\ddagger$ & left of, front of, left, top, gray, red, blue, green, cube, sphere, rubber & near, far, behind, right of, right, bottom, brown, cyan, purple, yellow, cylinder, metal & 7 & up to 3 & up to 3 & \checkmark \\
        \textsc{CityFlow-NL} & 1 & 6 & --- & 9 & Prog. & --- & Model & above, beneath, behind, suv, white, grey, van & to the left of, to the right of, in front of, sedan, black, red, blue, pickup truck & up to 4 & up to 3 & up to 3 & \checkmark \\
        \textsc{Charades} & 35 & 5 & 15$^\dagger$ & --- & Prog. & Model & --- & above, in front of, touching, leaning on$^\dagger$, wearing, drinking from, lying on$^\dagger$, writing on$^\dagger$, twisting$^\dagger$ & beneath, behind, in, holding, sitting on, standing on, covered by, carrying, eating, wiping$^\dagger$, have it on the back$^\dagger$ & up to 4 & up to 3 & up to 2 & x \\
        \hline
    \end{tabular}
    \label{table:datasets}
    \end{table*}
}

\newcommand{\queryTable}{
    \begin{table}[t]
    \centering
    \caption{Query templates.}
    \small
    \begin{tabular}{cccccc}
        \toprule
        Dataset & {\# queries} & {\# Test} & Attr & \makecell{Spatial \\ Rel} & \makecell{Semantic \\ Rel} \\
        \midrule
        \textsc{CLEVRER} & 5000 & 5000 \\
        \textsc{CityFlow-NL} & 824 & 824 \\
        \textsc{Charades} & 4800 & 4801 \\
        \bottomrule
    \end{tabular}
    \label{table:datasets}
    \end{table}
}

\newcommand{\costTable}{
    \begin{table}[t!]
        \centering
        \caption{\revision{LLM cost estimation for executing one query over CLEVRER's full test set, with three missing UDFs.}}
        \label{table:cost}
        {\renewcommand{\arraystretch}{1}
        \begin{tabular}{c|cc}
            \hline
             Method & GPT-4o & Llama 3.1 70B \\ \hline
            {\system} & \$0.4 & \$0.03--0.11 \\
            {Vanilla LLM} & \$207 & \$59 \\
            {LLM for concepts} & \$1504 & \$430 \\ \hline
        \end{tabular}}
    \end{table}
}

\newcommand{\nltodslTable}{
    \begin{table}[t!]
        \centering
        \caption{Natural language to DSL correctness.}
        \label{table:nl_to_dsl}
        {\renewcommand{\arraystretch}{1}
        \begin{tabular}{c|cc}
            \hline
             Dataset & F1 = 1 & F1 $\geq$ 0.98 \\ \hline
            \textsc{CLEVRER} & 79\% & 100\% \\
            \textsc{CityFlow-NL} & 92\% & 99\% \\
            \textsc{Charades} & 83\% & 97\% \\ \hline
        \end{tabular}}
    \end{table}
}

\newcommand{\proposingUdfsTable}{
    \begin{table}[t!]
        \centering
        \caption{Proposing UDFs. FPs are incorrectly proposed UDFs, and FNs are the missed UDFs that \system fails to propose.}
        \label{table:proposing_udfs}
        {\renewcommand{\arraystretch}{1}
        \begin{tabular}{c|cccc}
            \hline
            Dataset & \# New UDFs & \# proposed UDFs & \# FP & \# FN \\ \hline
            \textsc{CLEVRER} & 540 & 441 & 27 & 126 \\
            \textsc{CityFlow-NL} & 270 & 261 & 4 & 14 \\
            \textsc{Charades} & 270 & 229 & 19 & 60 \\ \hline
        \end{tabular}}
    \end{table}
}

\newcommand{\programUdfPerformanceTable}{
    \begin{table}[t!]
        \centering
        \caption{Program-based UDFs performance.}
        \label{table:program_udf_performance}
        {\renewcommand{\arraystretch}{1}
        \begin{tabular}{c|cc|cc}
            \hline
            \multirow{2}{*}{Dataset} & \multicolumn{2}{c|}{best $=$ ``program''} & \multicolumn{2}{c}{best $\neq$ ``program''} \\ \cline{2-5}
            & best UDFs & all UDFs & best UDFs & all UDFs \\ \hline
            \textsc{CLEVRER} & 0.995 & 0.203 & 0.664 & 0.024 \\
            \textsc{CityFlow-NL} & 1.000 & 0.831 & 0.349 & 0.203 \\
            \textsc{Charades} & 1.000 & 0.092 & 0.162 & 0.030 \\ \hline
        \end{tabular}}
    \end{table}
}

\newcommand{\programUdfTypesTable}{
    \begin{table}[t!]
        \centering
        \caption{Program-based UDF types.}
        \label{table:program_udf_types}
        \setlength{\tabcolsep}{3pt}
        {\renewcommand{\arraystretch}{1}
        \begin{tabular}{c|cccc|cccc}
            \hline
            \multirow{2}{*}{Dataset} & \multicolumn{4}{c|}{best $=$ ``program''} & \multicolumn{4}{c}{best $\neq$ ``program''} \\ \cline{2-9}
            & all & reuse & param & pixel & all & reuse & param & pixel \\ \hline
            \textsc{CLEVRER} & 195 & 52 & 59 & 40 & 37 & 8 & 14 & 12 \\
            \textsc{CityFlow-NL} & 83 & 25 & 14 & --- & 91 & 90 & 16 & --- \\
            \textsc{Charades} & 70 & 32 & 15 & --- & 75 & 33 & 28 & --- \\ \hline
        \end{tabular}}
    \end{table}
}

\newcommand{\modelUdfPerformanceTable}{
    \begin{table}[t!]
        \centering
        \caption{Model-based UDFs performance.}
        \label{table:model_udf_performance}
        {\renewcommand{\arraystretch}{1}
        \begin{tabular}{c|cc|cc}
            \hline
            \multirow{2}{*}{Dataset} & \multicolumn{2}{c|}{best $=$ ``model''} & \multicolumn{2}{c}{best $\neq$ ``model''} \\ \cline{2-5}
            & model & dummy & model & dummy \\ \hline
            \textsc{CLEVRER} & 0.889 & 0.519 & 0.556 & 0.662 \\
            \textsc{CityFlow-NL} & 0.687 & 0.289 & 0.735 & 0.664 \\
            \textsc{Charades} & 0.724 & 0.207 & 0.302 & 0.728 \\ \hline
        \end{tabular}}
    \end{table}
}

\newcommand{\modelUdfLabelingQualityTable}{
    \begin{table}[t!]
        \centering
        \caption{VLM labeling quality.}
        \label{table:model_udf_labeling_quality}
        {\renewcommand{\arraystretch}{1}%
        \begin{tabular}{cc||cc|cc}
            \hline
            \multicolumn{2}{c||}{CityFlow-NL} & \multicolumn{4}{c}{Charades} \\ \cline{1-6}
            suv & 0.843 & holding & 0.776 & sitting on & 0.897 \\
            white & 0.906 & standing on & 0.846 & covered by & 0.749 \\
            grey & 0.797 & carrying & 0.761 & eating & 0.737 \\
            van & 0.956 & wiping & 0.685 & touching & 0.632 \\
            sedan & 0.903 & leaning on & 0.816 & wearing & 0.854\\
            black & 0.819 & drinking from & 0.916 & lying on & 0.857\\
            red & 0.973 & writing on & 0.894 & above & 0.683 \\
            blue & 0.904 & in front of & 0.644 & beneath & 0.561 \\
            pickup truck & 0.956 & behind & 0.599 & in & 0.444\\
            \hline
        \end{tabular}}    
    \end{table}
}

\newcommand{\choosingUdfTypeTable}{
    \begin{table}[t!]
        \centering
        \caption{Number of correctly selected UDF types. The ``No.'' columns represents the number of proposed UDF instances.}
        \label{table:choosing_udf_type}
        {\renewcommand{\arraystretch}{1}
        \begin{tabular}{c|cccc|cc}
            \hline
            \multirow{2}{*}{Dataset} & \multicolumn{4}{c|}{best $\neq$ ``dummy''} & \multicolumn{2}{c}{best $=$ ``dummy''} \\ \cline{2-7}
            & No. &\texttt{both} & \texttt{llm} (\texttt{gpt-4o}) &  \texttt{llm} (\texttt{gpt-4-turbo}) & No. & \texttt{both} \\ \hline
            \textsc{CLEVRER} & 225 & 203 (90\%) & 208 (92\%) & 145 (64\%) & 7 & 3 (43\%) \\
            \textsc{CityFlow-NL} & 174 & 154 (89\%) & 170 (98\%) & 102 (59\%) & 0 & 0 (---)\\
            \textsc{Charades} & 125 & 90 (72\%) & 104 (83\%) & 57 (46\%) & 20 & 18 (90\%) \\ \hline
        \end{tabular}}
    \end{table}
}

\newcommand{\UdfSelectionTable}{
    \begin{table}[t!]
        \centering
        \caption{UDF selection performance \revision{and selected UDF distribution, using ``\texttt{both}'' strategy}.}
        \label{table:udf_selection}
        {\renewcommand{\arraystretch}{1}
        \begin{tabular}{c|ccc|ccc}
            \hline
            Dataset & No. & best & 80\% of best & program & model & dummy \\ \hline
            \textsc{CLEVRER} & 232 & 163 (70\%) & 218 (94\%) & 183 (79\%) & 37 (16\%) & 12 (5\%) \\
            \textsc{CityFlow-NL} & 174 & 144 (83\%) & 154 (89\%) & 94 (54\%) & 76 (44\%) & 4 (2\%)\\
            \textsc{Charades} & 145 & 93 (64\%) & 123 (85\%) & 81 (56\%) & 23 (16\%) & 41 (28\%) \\ \hline
        \end{tabular}}
    \end{table}
}

\newcommand{\IntentAmbiguityTable}{
    \begin{table}[t!]
    \centering
    \caption{Evaluated interpretations of ``behind''.}
    \label{table:intent_ambiguity}
    {\renewcommand{\arraystretch}{1}
    \begin{tabular}{p{0.13\columnwidth}|p{0.80\columnwidth}}
        \hline
        Concept & Interpretation \\ \hline
        behind-0 & If the center of o1 is above the center of o2. \\ \hline
        behind-1 & If the center of o1 is below the center of o2. \\ \hline
        behind-2 & If the bottom edge of o1 is above the bottom edge of o2. \\ \hline
    \end{tabular}}
    \end{table}
}

\section{Introduction}

Rapid advances in video analytics have fueled the development of innovative applications across various fields. In medical education, surgery videos enhance students' procedural knowledge by illustrating complex temporal and spatial events~\cite{FRIEDL20061760}. In biology, scientists use wildlife footage to study organism behaviors and interactions in their natural habitats~\cite{SCHLINING201396,dellinger:21}. In transportation, traffic surveillance systems analyze and manage traffic flow, improving urban mobility~\cite{DBLP:conf/intellisys/HammerLFLW20}. Across these applications, analysts seek to query video databases for events characterized by spatio-temporal and semantic interactions. For instance, an analyst might search for ``a motorcycle swerving near a silver Subaru and then colliding with it'' or ``a doctor holding a scalpel and then placing it on a table.''

\teaserFigure

Though promising, answering video queries using frontier vision-language models (VLMs) remains underwhelming.
Although VLMs have demonstrated notable capabilities on diverse, challenging tasks~\cite{DBLP:conf/icml/RadfordKHRGASAM21, DBLP:conf/nips/LiuLWL23a, DBLP:conf/icml/0008LSH23}, they struggle to answer \textit{compositional queries}~\cite{DBLP:conf/cvpr/MaHGGGK23} that involve
recognizing objects (e.g., ``car'', ``truck''), reasoning about relationships (e.g., ``behind'', ``holding''), and identifying attributes (e.g., ``silver color'', ``Subaru make''). 
This challenge is further amplified when queries require temporal reasoning~\cite{DBLP:conf/nips/Yu0YB23}
(e.g., ``X then Y'', ``X for at least 10 seconds''). While new models continue to improve their ability to reason spatially~\cite{DBLP:journals/corr/abs-2401-12168} and compositionally~\cite{hu2023visual}, their performance remains low~\cite{DBLP:journals/corr/abs-2406-14852}.
Additionally, deploying large models at scale is prohibitively expensive~\cite{qiao2024scaling} and inference is slow~\cite{wan2023efficient}. For example, current VLMs can only achieve a throughput of around 100 tokens per second~\cite{Qwen-VL, gpt_throughput}, making their use in large-scale video analytics intractable.

Alternatively, \textit{workflow-oriented} video data management systems (VDBMSs) answer compositional queries by decomposing them into granular subtasks~\cite{DBLP:conf/cvpr/AndreasRDK16, DBLP:conf/cvpr/GuptaK23, DBLP:conf/iccv/SurisMV23, DBLP:conf/nips/0001ST00Z23, DBLP:journals/corr/abs-1910-02993, DBLP:journals/pvldb/BastaniMM20, DBLP:journals/pvldb/ZhangD0HKB23, DBLP:conf/cav/MellBZB23}.
Submodules identify and track objects, attributes, and relationships across frames, forming spatio-temporal scene graphs~\cite{DBLP:journals/ijcv/KrishnaZGJHKCKL17,DBLP:conf/cvpr/JiK0N20}. These scene graphs can expressively represent many complex visual queries. Various scene graph generation techniques~\cite{DBLP:conf/cvpr/ZellersYTC18, DBLP:conf/cvpr/DaiZL17} have been proposed to extract scene graphs from images and videos. In these systems, subtasks that extract scene graph elements are solved individually and then composed to answer a compositional query~\cite{lu2016visual}.

However, VDBMSs make a critical assumption: the existence of modules capable of executing subtasks to answer a complex query. Systems typically provide a variety of built-in modules ~\cite{DBLP:conf/cvpr/GuptaK23, DBLP:conf/iccv/SurisMV23, DBLP:conf/nips/0001ST00Z23, DBLP:journals/pvldb/ZhangD0HKB23, DBLP:conf/cav/MellBZB23} and often allow extensibility via user-defined functions (UDFs) for unsupported scenarios~\cite{DBLP:journals/pvldb/ZhangD0HKB23, DBLP:conf/sigmod/XuKAR22, DBLP:journals/corr/abs-2403-14902}. Concerning our motorcycle query, a user might need to supply a UDF to filter for ``silver'' or nearby objects if the system lacks these capabilities. Despite the availability of pre-trained computer vision models that can be readily integrated as UDFs, users may require solutions for domain-specific applications or seek to identify fine-grained object classes and subjective concepts for which no off-the-shelf models exist. Identifying or adapting domain-specific models in such cases may be possible, though tedious. Additionally, it
may be necessary to---on a per-UDF basis---curate datasets and perform extensive training to achieve satisfactory performance~\cite{DBLP:conf/cvpr/AndreasRDK16}.

To address these challenges, we present \system (\Cref{fig:teaser}), a \textit{self-enhancing} VDBMS that empowers users to flexibly issue and answer compositional queries, even when the necessary modules are unavailable. To use \system, a user only needs to provide a video dataset and a natural language (NL) description of the query. \system then \textit{automatically} identifies and builds the necessary modules and encapsulates them as new UDFs to expand its querying capabilities. It then compiles the NL query into a declarative one that it executes over the video dataset.

There are several challenges in building \system. First, given an NL query, \system converts it into a declarative one that it can execute. Unfortunately, existing methods assume the existence of predefined modules that can be invoked to construct the declarative query~\cite{DBLP:journals/pvldb/ZhangD0HKB23,DBLP:conf/cidr/DaumZHBHKCW22}. \system addresses this challenge by identifying any semantic concepts that are not supported by existing modules or UDFs during compilation and leveraging an LLM's reasoning to determine when and which new UDFs are needed.

The second challenge involves transforming the various identified missing semantic concepts into executable modules. While LLMs can produce useful code in various contexts~\cite{DBLP:journals/corr/abs-2107-03374, DBLP:journals/corr/abs-2108-07732, DBLP:conf/nips/YeCDD23, chen2023seed}, queries on video data are often highly ambiguous and their performance varies significantly (see~\Cref{subsec:exp_udf_generation}). To address the challenge of handling a multitude of missing concepts, \system automatically implements two types of UDFs---\textit{program-based UDFs} and \textit{distilled-model UDFs}. In our system, program-based UDFs are imperative Python functions that operate on relational tables and video pixels. While this class of UDFs can classify many relationships and attributes with high quality~\cite{DBLP:journals/pvldb/ZhangD0HKB23, Chen2019SceneGP}, we show in~\Cref{sec:distilled_model_udfs} the need for distilled-model UDFs, which are lightweight machine learning (ML) models that are trained on the fly to classify more nuanced concepts~\cite{toubal2024modeling}. \system leverages LLMs for both UDF types: program-based UDFs harness an LLM's programming capability, whereas distilled-model UDFs rely on the LLM's ability to annotate and convert visual concepts into ML models~\cite{toubal2024modeling}. To mitigate potential erroneous LLM-generated UDFs, \system implements syntax and semantic verification steps. 

The third challenge involves managing the inherent ambiguity in
user intent when articulating a query. This is especially important for specialized and subjective concepts that are difficult to resolve without user feedback. For example, users may have different interpretations of the ``near'' relationship. 
Further compounding this difficulty is our observation that for some inputs a program-based UDF might perform better than a distilled-model UDF, while in other cases the opposite is true.
Finally, 
for the most challenging semantic concepts, \textit{no} UDF may perform well and employing one might risk overall query performance. 
To address this formidable combination of challenges, \system generates \textit{multiple} candidate UDFs each with different semantic interpretations, implementations, and properties. \system then employs active learning to efficiently identify the variant that best matches the user's intent.

Finally, the self-building nature of our system necessitates a consistent and unified UDF model. Typically, a UDF can be an arbitrary function that operates on database tuples. However, one challenge lies in modeling UDFs in a structured way to handle concepts of objects, relationships, and attributes while seamlessly integrating and interacting with various system components. To address this, we propose a unified UDF scheme for different semantic concepts, which enables the LLM to generate UDFs in a structured format and simplifies the compilation process. By incrementally growing the database, future UDFs can be composed using existing ones.

We evaluate \system on three video datasets from different domains~\cite{CLEVRER2020ICLR, DBLP:journals/corr/abs-2101-04741, DBLP:conf/cvpr/JiK0N20} and show that it significantly improves query performance, in terms of F1 score, by automatically selecting, implementing, and executing its automatically-generated UDFs. We finally conduct a thorough analysis of \system to understand its efficiency in terms of execution time and monetary cost.

Overall, \system's self-enhancing capability is an important step toward making VDBMSs more practical to deploy and use in a variety of applications.

\section{Background}~\label{sec:problem_setup}

\system leverages EQUI-VOCAL's~\cite{DBLP:journals/pvldb/ZhangD0HKB23} scene graph data model and query language, which models  compositional video events as spatio-temporal scene graphs.  This approach draws from cognitive foundations in human perception~\cite{zacks2001perceiving,kurby2008segmentation,biederman1987recognition} and enables a variety of compositional queries~\cite{DBLP:journals/pvldb/ZhangD0HKB23}.
This section summarizes key background information about these concepts.

\relationalSchema

\textbf{Data model.}
In \system, each video comprises a series of $N$ frames $\{f_1, \ldots, f_N\}$. The visual content of each frame is represented by a \textit{scene graph} $g_i = (\textbf{o}_i, \textbf{r}_i)$, capturing all \textit{objects} $\textbf{o}_i$ and all \textit{relationships} $\textbf{r}_i$ between those objects within the frame at some time. Objects may also possess \textit{attributes}.
While a relationship links two objects, an attribute is attached to one object.
A \textit{region graph} $g_{ij}$ is a subgraph of $g_i$, i.e., $g_{ij} \subseteq g_i$, that contains information critical for identifying an event.
Finally, an \textit{event} $e$ is a sequence of region graphs $ e = \{g_1, \ldots, g_{k}\} $, where region graphs with a smaller index occur earlier in time than those with a larger index, but they do not need to be contiguous or distinct.
The relational schema in~\Cref{table:relational_schema} captures the scene graphs data model.

\textbf{Query language.}
In \system, a query identifies video segments that match a user-specified event. \system supports relational queries over the schema in~\Cref{table:relational_schema}, which are of the following form. Using Datalog notation: $ q(vid) \textrm{ :- } g_1, \ldots, g_k, \textbf{p}, \textbf{d}, w $, where
$g_1, \ldots, g_k$ is a temporally ordered sequence of region graphs specifying that a matching event consists of $g_1$, followed by $g_2$, followed by $g_3$, etc. Each $g_i$ can persist for multiple frames and there can be other frames between $g_i$ and $g_{i+1}$. $\textbf{p}$ is a set of predicates that are applied to objects, relationships, and attributes in region graphs. $\textbf{d}$ is a set of constraints on the duration for which a region graph must remain valid
before transitioning to the next one. Lastly, $w$ is the maximum number of frames between $g_1$ and $g_{k}$. As an example, the event ``A car is initially far from a truck, then remains close to the truck for more than 10 seconds'' can be expressed as in~\Cref{lst:datalog_snippet}.

\begin{lstlisting}[
    float=t!,
    language=Prolog, 
    basicstyle=\small\ttfamily,
    caption={An example of the query language.},
    label={lst:datalog_snippet},
    aboveskip=0pt, %
    belowskip=-1em, %
    lineskip=0pt %
]
  g1(vid, fid, fid, oid1, oid2) :- Objects(vid, fid, oid1, 'car', _, _, _, _), 
      Objects(vid, fid, oid2, 'truck', _, _, _, _), 
      Relationships(vid, fid, _, oid1, 'far', oid2), oid1 != oid2.
  g2(vid, fid, oid1, oid2) :- Relationships(vid, fid, _, oid1, 'near', oid2), oid1 != oid2.
  g2_star(vid, fid, fid, oid1, oid2) :- g2(vid, fid, oid1, oid2).
  g2_star(vid, fid_start, fid_end, oid1, oid2) :- g2_star(vid, fid_start, fid, oid1, oid2), 
      g2(vid, fid_end, oid1, oid2), fid_end = fid + 1.
  q(vid) :- g1(vid, fid11, fid12, oid1, oid2), g2_star(vid, fid21, fid22, oid1, oid2), 
      fid21 > fid12, fid22 - fid21 + 1 > 10 * 24. %
\end{lstlisting}

\textbf{DSL.}
\system adopts EQUI-VOCAL's domain-specific language (DSL)~\cite{DBLP:journals/pvldb/ZhangD0HKB23}, which encapsulates query logic while abstracting away SQL details. The query executor compiles these DSL queries into efficient SQL for execution over relational tables. Using the DSL, the same event can be expressed as (assuming 24 frames per second):
$(\texttt{Car}(o_1), \texttt{truck}(o_2), \texttt{far}(o_1,o_2); \texttt{Duration}(\texttt{near}(o_1,o_2), 240)$.
In this DSL, the \textit{variable} $o$ represents an arbitrary object in a query, with distinct subscripts indicating objects with different $oid$'s. All predicates in a region graph are separated by commas. Region graphs are then sequenced in temporal order using semicolons. Each region graph can persist for multiple frames and there can be other frames between two adjacent region graphs. Finally, $\texttt{Duration}(g, d)$ stipulates that the region graph $g$ exists in \textit{at least} $d$ consecutive frames. The query returns a set of video segment identifiers.
\section{A UDF-based data model}~\label{sec:udf_data_model}

In this section, we formalize the types of UDFs supported by \system.
These UDFs allow users to define custom objects, relationships, and attributes.
We then describe how UDFs are compiled and executed in \system.

\Cref{table:relational_schema} shows \system's relational schema. The \texttt{Frames} relation includes a virtual \texttt{pixels} column that stores the frame pixel values in a 3D array  ($H \times W \times 3$, where $H$ is height, $W$ is width, and 3 represents the color channels). The \texttt{Objects}, \texttt{Relationships}, and \texttt{Attributes} relations store corresponding detected frame elements.

\begin{lstlisting}[
    float=t!,
    language=SQL, 
    basicstyle=\small\ttfamily,
    caption={SQL query to identify frames where a silver car is behind a truck.},
    label={lst:sql_snippet},
    aboveskip=0pt, %
    belowskip=-1em, %
    lineskip=-1pt %
]
SELECT DISTINCT f.vid, f.fid, o1.oid, o2.oid
FROM frames f, objects AS o1, objects AS o2
WHERE f.vid = o1.vid AND f.fid = o1.fid AND o1.vid = o2.vid 
    AND o1.fid = o2.fid AND o1.oid <> o2.oid 
    AND car(o1.oname) = TRUE AND truck(o2.oname) = TRUE 
    AND behind(o1.y1, o1.y2, o2.y1, o2.y2) = TRUE 
    AND silver(f.pixels, o1.x1, o1.y1, o1.x2, o1.y2) = TRUE
\end{lstlisting}

A UDF extends database functionality. A typical UDF takes columns as input, returns a scalar value or a row set, and is used in SQL statements, e.g., in \texttt{WHERE} clauses. When querying video databases, \system supports generating and executing UDFs to identify custom objects, relationships, and attributes, enabling users to find complex, compositional events. \Cref{lst:sql_snippet} shows an example SQL query with UDFs for the objects \texttt{car} and \texttt{truck}, the relationship \texttt{behind}, and the attribute \texttt{silver}. \system supports imperative and declarative Python-based UDFs, and are categorized into the following classes: relationship, attribute, object, and value-lookup.

A \textbf{relationship UDF} or \textbf{attribute UDF} is a deterministic, scalar predicate that indicates whether an input exhibits a specified relationship or attribute. It accepts zero (i.e., a dummy UDF; see~\Cref{subsec:dummy_udf}) or more columns as arguments and produces a boolean result. The parameters can include any of the following columns from each table in the SQL query's \texttt{FROM} clause: \texttt{pixels}, \texttt{oname}, \texttt{x1}, \texttt{y1}, \texttt{x2}, \texttt{y2}, \texttt{rname}, and \texttt{aname}.
\system restricts relationship UDFs and attribute UDFs to be frame-level, i.e., they operate on object(s) within the same frame. Therefore, input arguments 
are all from the same video frame (identified by \texttt{vid} and \texttt{fid}) and are associated with one or two
distinct objects.
As an example, we might define a relationship UDF to indicate whether object $o_1$ is behind another object $o_2$ by comparing their centroid $y$-coordinates:
\begin{minted}[fontsize=\small]{python}
def behind(o1_y1, o1_y2, o2_y1, o2_y2):
    return (o1_y1 + o1_y2) / 2 < (o2_y1 + o2_y2) / 2
\end{minted}
We could also define an attribute UDF to indicate if a detected car is silver by running an ML model over the frame pixels:
\begin{minted}[fontsize=\small]{python}
def silver(pixels, x1, y1, x2, y2):
    cropped_img = pixels[y1:y2, x1:x2]
    is_silver = awesome_color_classifier(cropped_img)
    return is_silver
\end{minted}

An \textbf{object UDF} requires localizing, classifying, and tracking objects in videos. Instead of returning a boolean value, it is a table-valued function that takes a video segment, \texttt{video\_pixels}, as input. A video segment comprises frames with the same \texttt{vid} concatenated into a 4D array with an additional dimension for the frame index. An object UDF detects and tracks objects of a specific class. Given a video segment, the UDF makes calls to a custom object detection and tracking model and returns a row set, which follows the \texttt{Objects} schema listing the detected and tracked objects.
As an example, we can define an object UDF that detects all cars in a video:
\begin{minted}[fontsize=\small]{python}
def car(video_pixels):
    obj_tuples = []
    car_detector, tracker = load_models()
    for frame in video_pixels: 
        detected_cars = car_detector(frame)
        tracked_objs = tracker.update(detected_cars)
        obj_tuples.extend(tracked_objs)
    return obj_tuples
\end{minted}

In this paper, we assume that object UDFs are given and focus on proposing and generating relationship and attribute UDFs. We leave the extension to object UDFs for future work.

A \textbf{value-lookup UDF} is a class of UDFs that simply encapsulate a predicate over existing column values.
As an example, suppose the value ``car'' is in the domain of the object \texttt{oname}. Then, a value-lookup UDF can be defined as:
\begin{minted}[fontsize=\small]{python}
def car(oname):
    return oname == 'car'
\end{minted}

While value-lookup UDFs are not strictly necessary, as they can be directly and easily expressed in SQL statements, we wrap all predicates of our DSL queries in UDFs to simplify the compilation from the DSL to SQL.

\relationalView

UDFs can have different lists of parameters in their signatures.  
To execute a query with UDFs, \system first constructs the relational views as shown in~\Cref{table:relational_view} derived from the relations in~\Cref{table:relational_schema}. The views--\texttt{ObjView}, \texttt{RelView} and \texttt{AttrView}--contain all attributes that an object UDF, relationship UDF, and attribute UDF can potentially accept as input arguments, respectively. For simplicity, \system generates UDFs with a uniform list of parameters for each class. 

Object, relationship, and attribute UDFs can be expensive to evaluate, as they may operate on image pixels and invoke ML models. To optimize query execution, \system caches the results of all
UDFs and replaces them with value-lookup UDFs. When executing a UDF over a video corpus for the first time, \system materializes the results to make it available as a value in the corresponding column and substitutes the UDF with a value-lookup UDF of the same name.
For example, when running a query with the \texttt{silver} attribute UDF for the first time, \system evaluates the UDF for each object in the video corpus. If an object \texttt{OID1} in a frame \texttt{F1} of a video segment \texttt{V1} is classified as silver, \system inserts a new row with values (\texttt{V1}, \texttt{F1}, \texttt{OID1}, `silver') into the \texttt{Attributes} relation. Then, \system replaces the \texttt{silver} attribute UDF with a value-lookup UDF  that checks whether the \texttt{aname} is `silver'. 
Reusing the results of predicate evaluation for query optimization is a long-standing research topic~\cite{DBLP:conf/sigmod/XuKAR22, DBLP:conf/sigmod/RoySSB00, DBLP:conf/sigmod/MistryRSR01, DBLP:journals/pvldb/ElghandourA12} and is not the focus of this paper. Batch inference can further accelerate ML-based UDFs.
\section{\system approach}~\label{sec:generating_udf}

\systemFigure

\system needs to address several challenges. First, it must determine whether existing UDFs can adequately answer a user query, or if new UDFs should be created (C1). Second, \system should support implementing UDFs drawn from diverse range of semantic concepts (C2). Third, since \system utilizes error-prone LLMs to generate UDFs, it is crucial to ensure high quality in the produced UDFs (C3). We now discuss our solutions to each challenge.

\Cref{fig:system} shows the architecture of \system
. The user initializes the system with a video dataset and an optional set of UDFs. After preprocessing (\Cref{sec:implementation}), the user can issue NL queries to identify compositional events within the videos. The \textit{Query Parser} (\Cref{sec:proposing_udfs}) parses the query into the DSL notation described in~\Cref{sec:problem_setup}. If successful, the DSL query is passed to the \textit{Query Executor} to find all matching videos in the dataset. If the query contains predicates that existing UDFs cannot resolve, the \textit{UDF Proposer} (\Cref{sec:proposing_udfs}) is invoked to propose the names and descriptions of new UDFs. The \textit{UDF Generator} (\Cref{sec:program_based_udfs,sec:distilled_model_udfs}) creates executable candidates for each proposed UDF. Next, the \textit{UDF Selector} (\Cref{sec:udf_selection}) solicits user labels to select the implementation that best aligns with the user's intent. 
With updated UDFs, the \textit{Query Parser} re-parses the query for execution. The \textit{Storage Manager} maintains the raw videos, extracted relational information, and available UDFs.

\subsection{Query parsing and UDF proposal}~\label{sec:proposing_udfs}

The Query Parser converts NL queries into our DSL and determines the need for new UDFs. 
\system must understand the semantics of the user's query and available UDFs, mapping each part of the query to an existing UDF or suggesting the creation of a new one. 
Moreover, \system needs to be resilient to the linguistic ambiguities and synonymous terms in NL queries.

\system utilizes LLMs to convert complex NL to a consistent DSL format. LLMs show strong capabilities in SQL and program generations~\cite{DBLP:conf/nips/PourrezaR23, DBLP:conf/nips/WangW0CSK23, DBLP:conf/cvpr/GuptaK23} via in-context learning without fine-tuning. We incorporate domain-specific constraints into LLM prompts to ensure adherence to DSL grammar and conduct post-verification to ensure syntax correctness. 
Inspired by Wang et. al.~\cite{DBLP:conf/nips/WangW0CSK23} and Hsieh et. al.~\cite{Hsieh2023ToolDE},
\system provides the DSL definition, UDF format, and descriptions of available UDFs as NL documentation. This enables \system to generate grammatically correct DSL queries, determine whether new UDFs need to be created, and interact with LLMs in a zero-shot manner (C1). 
\revision{
While \system focuses on zero-shot prompting, introducing few-shot examples~\cite{DBLP:conf/nips/BrownMRSKDNSSAA20} could potentially further improve UDF proposal performance. However, 
referencing unavailable UDFs in examples may lead LLMs to incorrectly assume their availability at query time. 
Since the available UDFs can change as the database evolves, examples must be carefully curated for different database states. Figuring out the optimal prompting strategy, though beneficial, is not the focus of this paper. 
}
\iftechreport
An example prompt is shown in~\Cref{fig:parsing_query_prompt,fig:proposing_udf_prompt}.
\fi 

To enhance the reliability of the LLM's response, \system performs post-verification of the generated DSL query to ensure it is syntactically valid and only uses available UDFs.
If parse errors are encountered, \system appends the error message to the context and asks the LLM to make another attempt.

When the query contains predicates that cannot be resolved using the available UDFs, we prompt the LLM to identify this
and propose new UDFs. The output from this process is the function signature and textual description of the proposed UDF generated by the LLM. In case of a query with multiple missing UDFs, \system returns a list of proposed UDFs and will generate them one by one. 
An example proposed UDF of \texttt{behind} is:  

\begin{lstlisting}[
    float=h,
    language=python, 
    basicstyle=\small\ttfamily,
    label={lst:proposed_udf},
    aboveskip=0pt, %
    belowskip=-0.5em, %
    lineskip=-1pt %
]
{"signature": "behind(o0, o1)", "description": "Whether o0 is behind o1"}
\end{lstlisting}

Since both the Query Parser and UDF Proposer utilize LLMs, their outputs may sometimes be inaccurate. If the LLM proposes fewer UDFs than needed, it may make performance improvements less noticeable but will not degrade performance. Conversely, proposing more UDFs can increase system runtime and monetary costs; however, as detailed in~\Cref{sec:udf_selection}, we have implemented techniques to ensure that generated UDFs (whether superfluous or inaccurate) do not adversely affect performance. Finally, users may also rephrase their NL query if they are not fully satisfied with the query results.

We show the effectiveness of our approach empirically in~\Cref{sec:evaluation} for queries with detailed and explicit descriptions. 
\revision{Semantic parsing (e.g., text-to-SQL) is a highly active research area with significant recent advances~\cite{DBLP:journals/vldb/KatsogiannisMeimarakisK23, DBLP:conf/coling/Deng0022, DBLP:journals/corr/abs-1709-00103, DBLP:conf/emnlp/YuZYYWLMLYRZR18, DBLP:journals/pvldb/GaoWLSQDZ24}. While current methods still fall short of human-level performance~\cite{DBLP:conf/nips/LiHQYLLWQGHZ0LC23}, improving text-to-SQL is not the goal of our paper. \system focuses on the efficient generation of missing UDFs, relies on LLMs currently adequate performance for text-to-SQL in our context, and will benefit from future advances in this area.}

\subsection{Program-based UDF generation}~\label{sec:program_based_udfs}

The UDF Generator implements executable UDFs based on the LLM-proposed UDF signatures and descriptions. As discussed, \system should produce high-quality UDFs (C3) for a wide range of semantic concepts (C2). While \system leverages the programming capabilities of LLMs~\cite{DBLP:journals/corr/abs-2308-12950, DBLP:journals/corr/abs-2303-12712, chen2023teaching, chen2022codet} to generate UDFs as Python programs, more work is required. Some tasks require complex visual understanding, rendering it challenging to solve via programs (e.g., determining a car's make). Additionally, even the most advanced LLMs remain error-prone and their programming performance for nuanced tasks is not yet on par with humans.

To address the first problem, \system supports two types of UDF implementations: program-based UDFs, which are Python programs generated by LLMs (\Cref{sec:program_based_udfs}), and distilled-model UDFs, which are lightweight vision models distilled from strong pretrained models (\Cref{sec:distilled_model_udfs}). To improve generated UDF quality, \system uses a two-step approach by verifying both syntactic (\Cref{subsec:syntax_verification}) and semantic (\Cref{sec:udf_selection}) correctness. 

At a high level, given a UDF signature $h$ and description $d$, a video database instance $I$ over schema $R$, \system performs the following steps to generate a UDF $p$:
\begin{enumerate}[label=\arabic*., leftmargin=*]
    \item \textit{Generate UDF candidates using an LLM}.  Given $h$, $d$, $I$, and $R$, \system prompts an LLM to generate a set of $k$ candidate Python functions $\{p_1, ..., p_k\}$.
    \item \textit{Syntactically verify candidates}. \system executes each candidate program $p_i$ on a small dataset sampled from $I$ to verify syntactic correctness, ensuring that all passing candidates are executable. 
    \item \textit{Semantically verify candidates}. \system finally evaluates the semantic correctness of the remaining candidates and selects the best one (see \Cref{sec:udf_selection}).
\end{enumerate}
We next describe each step in further detail.

\subsubsection{Generating candidate programs using an LLM}~\label{subsec:program_based_udfs}

\programCandidatesFigure

In its most basic form, \system prompts the LLM to generate a program $p$ based on the UDF signature $h$ and description $d$. To ensure $p$ can be eventually expressed as a SQL predicate, \system rewrites $h$ in the DSL format to $h'$ that accepts columns from the schema $R$ as inputs. For example, the \texttt{behind} function is rewritten from \texttt{behind(o0, o1)} to \texttt{behind(o1\_y1, o1\_y2, o2\_y1, o2\_y2)}, where \texttt{o1\_y1}, \texttt{o1\_y2}, \texttt{o2\_y1}, \texttt{o2\_y2} are columns of $R$. The generated Python program's input is a set of attribute values that correspond to one or two objects in the video database, depending on the number of variables in $h$. It generates code that operates on those values, including the \texttt{pixels} column, and returns a boolean indicating if the predicate is satisfied or not. \Cref{fig:program_candidates} (program $p_1$) shows an example program UDF for the \texttt{behind}($\cdot$) predicate. 

To generate quality UDFs, \system must resolve three challenges. First, NL descriptions are often ambiguous and may not entirely capture the user's intent~\cite{lahiri2023interactive} (e.g., user-specific definition of ``far''). Second, predicates in video compositional queries often include hyperparameters that need to be tuned~\cite{DBLP:conf/cav/MellBZB23, DBLP:journals/pacmpl/MellZB24} for different datasets and user intents, e.g., determining the threshold distance for the ``far'' relationship. Lastly, as \system expands its collection of UDFs and database incrementally, it is essential that new UDFs be able to utilize results from previously established ones.

To address the challenge of linguistic ambiguity, we provide the LLM with $h'$ and $d$ and ask it to generate a list of $k$ candidate programs with a variety of semantic interpretations: $C=\{(p_1, s_1), \cdots, (p_k, s_k)\}$, where $p_i$ and $s_i$ respectively denote a Python program and its semantic interpretation. 
\iftechreport
\Cref{fig:implementing_udf_prompt} shows an example prompt used by \system in this step. 
\fi 
Later, \system verifies and selects one program for each proposed UDF. See \Cref{fig:program_candidates} for candidate program examples. 

To resolve the challenge of parametric predicates, \system additionally prompts the LLM to generate UDFs with optional numeric hyperparameters and their valid ranges.
More formally, \system prompts the LLM to produce
a (possibly empty) list $\Theta=\{(\theta_1, \textrm{df}_{\theta_1}, \textrm{min}_{\theta_1}, \textrm{max}_{\theta_1}), \cdots, (\theta_i, \textrm{df}_{\theta_k}, \textrm{min}_{\theta_k}, \textrm{max}_{\theta_k})\}$ of parameter names $\theta_i$, default value ${df}_{\theta_i}$, and range $[\textrm{min}_{\theta_i}, \textrm{max}_{\theta_i}]$.
In later steps, \system instantiates each hyperparameter with its default value as well as values sampled from the range and then selects the best program. See \Cref{fig:program_candidates}\textcircled{\small 2} for an example of a program that relies on a hyperparameter.

To resolve the final challenge of incrementally building the video database, our
approach is to additionally include in the LLM prompt, the active domain of all attributes populated by UDFs, which include \texttt{oname}, \texttt{rname}, and \texttt{aname}. This additional information enables the LLM to leverage \revision{the results of} existing UDFs as building blocks to dynamically compose more complex UDFs. \Cref{fig:program_candidates}\textcircled{\small 3} shows an example program inlining an existing UDF.

Overall, program-based UDFs are well-suited for concepts involving bounding box-like spatial relationships. In addition, other attributes and the \texttt{pixels} column can also be used to reason about existing concepts and perform statistical analysis of the pixels in a frame. A program-based UDF could also invoke a pretrained model over each frame, but this is in general expensive and slow. As discussed in~\Cref{sec:distilled_model_udfs}, a distilled-model UDF is a more efficient approach for concepts that require visual understanding of videos.

\subsubsection{Syntax verification}~\label{subsec:syntax_verification}

Prior works have proposed various approaches to improve the performance of LLM-generated programs. One line of work leverages unit tests to verify the functional correctness of generated programs~\cite{DBLP:journals/corr/abs-2305-17126, hu2023visual, Chen2023GENOMEGN, DBLP:journals/corr/abs-2107-03374}. This approach is not suitable for \system because users do not know in advance what UDFs will be generates, and thus cannot provide labeled data before issuing a query. Another line of work leverages LLMs to automatically generate test cases~\cite{chen2022codet}, evaluate the generated programs~\cite{chen2023teaching, DBLP:conf/nips/ShinnCGNY23}, and select the best one~\cite{DBLP:conf/iclr/0002WSLCNCZ23}. However, the ambiguity of semantic concepts in video queries means that the correct program is not always unique and often many of the generated programs are reasonable. As a result, the best program cannot be easily identified without user feedback and a working dataset.

Additionally, the quality of an LLM is not guaranteed. To address this issue, 
\system uses a two-step approach to verify and select the best program: syntax verification and semantic verification. In the first step, \system focuses solely on the syntax correctness of the programs. \system executes each candidate program on a small sample of data from the database $I$ and checks whether: (i) the number and types of inputs and outputs are correct, (ii) the program can be executed with the data samples, and (iii) $\Theta$, if any, can be parsed successfully.
If the verification fails, \system appends the error message to the context and prompt the LLM to make another attempt. 
\system draws samples by constructing tuples that contain attributes and values that respectively correspond to one or two objects in the database.
If the program still fails after a few trials, \system discards the program.
In our prototype, we empirically set the number of trials to five. 
\revision{Recent research has demonstrated the potential for formally verifying the correctness of generated UDFs based on a given semantic interpretation~\cite{DBLP:journals/corr/abs-2410-15756}, which could be incorporated into \system to further enhance the reliability of its generated Python programs.}
After generating multiple candidate programs, \system next utilizes user labels to select the one that best aligns with user intent, which we discuss in~\Cref{sec:udf_selection}.

\subsection{Distilled-model UDF generation}~\label{sec:distilled_model_udfs}

While program-based UDFs are powerful and flexible for predicates that reason about existing concepts, bounding box coordinates, or perform a simple statistical analysis of pixel values, they struggle with tasks that require understanding the visual contents of frames. 
Even though pretrained models like VLMs can be used in a program-based UDF to classify relationships and attributes in a zero-shot manner, running such models over the entire video dataset is expensive. 
\revision{
For instance, applying GPT-4o to every frame of the 10,000 five-second videos in the CLEVRER dataset~\cite{CLEVRER2020ICLR} to evaluate a predicate (e.g., ``color-red'' or some more complex predicate) would cost approximately \$1,413. This cost remains high even when applying techniques such as downsampling or predicate push-down: There are about 1.28M video frames and 5.44M objects to consider. Even if 99\% of the image patches corresponding to these objects were filtered, the cost of evaluating a predicate on the remaining 1\% of objects would still be around \$14---far higher than generating one UDF (about \$0.15; see~\Cref{subsec:exp_end_to_end}).
}
\system's goal is to generate cheaper UDFs with visual understanding capabilities. 
However, training a lightweight image classifier from scratch would require the user to spend a lot of time labeling for just one concept. 

Model distillation is a common technique in machine learning to transfer knowledge from a large model to a smaller, more efficient model~\cite{DBLP:conf/acl/HsiehLYNFRKLP23, toubal2024modeling, DBLP:conf/kdd/BucilaCN06, DBLP:conf/nips/ChenCYHC17, DBLP:journals/corr/HintonVD15}. Modeling Collaborator~\cite{toubal2024modeling} is a newly proposed framework that leverages foundation models to train image classifiers for visual concepts using minimal user effort.  To do so, %
given a target concept and description, the system (i) mines relevant images from the public domain, (ii) uses foundation models to annotate sampled images, (iii) trains a lightweight classifier using features extracted from a pretrained model (e.g., CLIP) and labels annotated by the foundation models, and (iv) performs multiple rounds of active learning to further improve its performance.  

In \system, we adopt a similar approach to automatically construct lightweight image classifiers for new concepts without requiring user labeling but with the modifications needed to resolve three unique challenges presented in our compositional query setting. 
First, random sampling of the user dataset might not give enough positive samples for training, especially for rare concepts.
Second, \system generates UDFs for relationships and attributes, which differ from the concepts in~\cite{toubal2024modeling}. An image usually includes multiple objects, and \system needs a different prompting strategy to guide a VLM in classifying specific objects or pairs of objects. Finally, for the same reason, merely extracting features from the entire image is insufficient to train a good classifier. 

\dataLabelingFigure

\subsubsection{Image sampling}
To generate a distilled-model UDF, \system initially randomly samples frames from the user's video dataset for annotation. However, when the target visual concept is infrequent in the dataset, random sampling does not effectively collect enough positive samples for training. For instance, only 0.83\% human-object pairs have an ``eating'' relationship in the Charades~\cite{DBLP:conf/cvpr/JiK0N20} dataset. Our solution is to apply \textit{object-aware sampling} to bootstrap the sampling process. Since all objects of interest are already detected and tracked in the dataset, \system can filter out irrelevant objects and sample only those likely to be involved in the target concept. For example, when labeling the ``eating'' relationship, object classes like ``food'' and ``person'' are more relevant than ``car'' and ``window''. To do this, \system first asks an LLM for relevant object classes, and then only samples objects belonging to these classes. 
\iftechreport
\Cref{fig:filtering_objects_prompt} shows an example prompt of object-aware sampling.
\fi

\subsubsection{Data labeling}~\label{subsec:data_labeling}
\system uses VLMs to automatically label sampled video frames as positive or negative based on a UDF description. A VLM takes as input an image-text pair and outputs a textual response. However, using a video frame and the UDF description as a direct query is ineffective, since the concepts we are interested in target specific objects or pairs of objects, rather than the frame as a whole. Thus,
\system applies the following prompting strategy, with the goal of encouraging the VLM to focus on particular objects or interactions between two objects. 
For attribute concepts, we use the UDF description proposed in~\Cref{sec:proposing_udfs} as the text input, and create the image input by sampling an object from the \texttt{Objects} relation and cropping the video frame to include only the object.
For relationship concepts, \system augments the text input  with the class names and bounding box coordinates of the relevant objects in the video frame to provide more context to the VLM. It then generates an image patch cropped from the video frame that includes a pair of objects in the same frame from the \texttt{Objects} relation. \system further augments the image patch by overlaying a red box around the subject and a blue box around the target, thereby providing the VLM with directional information about the relationship. \Cref{fig:data_labeling} shows an example prompt for labeling the \texttt{behind} relationship.

\subsubsection{Model training}~\label{subsec:model_training}
Similar to~\cite{DBLP:conf/iccv/StretcuVHVFTZAL23, toubal2024modeling, DBLP:journals/pvldb/DaumZ0MHKB23}, \system leverages a pretrained vision model (e.g., CLIP) as the feature extractor and uses the feature-label pairs to train a multi-layer perceptron (MLP).
However, for relationship classification, directly extracting features from the image patch containing two objects would perform poorly, as the feature extractor is not aware of the object locations and cannot capture directional relationships (e.g., $o_1$ holding $o_2$ vs. $o_2$ holding $o_1$). To address this, \system concatenates features from three versions of an image patch: the original one containing both objects, one where everything except the subject is masked, and one where everything except the target is masked. For attribute UDFs, \system simply extracts features from the image patch containing the object. In addition, \system incorporates text features of object class names, derived from the \texttt{Objects} relation, to enhance the MLP's performance.
After training an initial model, \system uses an active learning approach similar to~\cite{DBLP:conf/iccv/StretcuVHVFTZAL23, toubal2024modeling} to iteratively improve the model. During each iteration, the trained MLP is run over the unlabeled dataset, \system selects a batch of samples with the highest uncertainty for labeling by the VLM. 
\system then retrains the MLP with the updated labeled dataset. 

\revision{
\system uses VLMs to automatically label data and the cost is bounded by the training data size, which is at most 500 frames in our evaluation. This is only 0.04\% of the 1.28M frames in the entire database for CLEVRER, 0.8\% of the 65K frames for CityFlow-NL, and 0.2\% of the 289K frames for Charades (see~\Cref{sec:evaluation}), which correspond to extremely low sampling rates. 
This cost is thus far lower than directly applying an LLM to the entire video dataset, even with downsampling.
} 
Although \system's distilled model process may occasionally deliver unsatisfactory performance, its UDF Selector component, described next, mitigates this by selecting the best candidate UDFs and filtering low-performing models.

\subsection{UDF Selection}~\label{sec:udf_selection}

For a given UDF, \system can generate both a distilled-model and a set of program-based implementations. When there are multiple implementation candidates, \system needs to select the best one. 
\system does not have any initial data to validate semantic correctness, only the user NL query. If \system were to collect labeled data for validation before query execution, the user would need to manually go through the video dataset to find positive and negative examples. Instead, \system utilizes active learning at query time to strategically request binary labels on carefully selected samples that are most likely to resolve disagreements among UDF candidates. 
\iftechreport
Additionally, \system leverages LLMs to generate the UDF Python code on behalf of the user (\Cref{sec:program_based_udfs}) to reduce user effort. Manually writing code snippets for a UDF would require knowledge of Python and could be a tedious trial-and-error process for the user, especially when a UDF requires hyperparameters, operates on pixel values, or leverages existing UDFs as building blocks. We now describe the UDF selection process in detail.
\fi

\begin{algorithm}[t!]
    \SetNoFillComment
    \SetKwInOut{Input}{Input}
    \SetKwInOut{Output}{Output}

    \Input{Set of unlabeled data $U$, set of UDF candidates $C$, and hyperparameters $b$, $n_s$, $t_p$, $t_n$
    }
    \Output{Selected UDF with highest score}
    $L_p \gets \{\}$, $L_n \gets \{\}$ \\
    $W \gets \{ w_i \mid w_i = 1 / |C|, i = 1, 2, \dots, |C| \}$ \label{line:W1} \\ 
    \For{$i = 1$ \textbf{to} $b$}
    {   
        $U_s \gets \textsc{SampleSubset}(U, n_s)$ \label{line:sample} \\ 
        \tcc{Phase 1: Bootstrapping}
        \If {$|L_p| < t_p$} { \label{line:bootstrap_start}
            $L_p', L_n' \gets \textsc{PickPositive}(U_s, C, W)$
        }
        \ElseIf{$|L_n| < t_n$} {
            $L_p', L_n' \gets \textsc{PickNegative}(U_s, C, W)$ \label{line:bootstrap_end} 
        }
        \tcc{Phase 2: Active learning}
        \ElseIf{$|L_p| < |L_n|$} { \label{line:active_learning_start} 
            $L_p', L_n' \gets \textsc{PickPositive}(U_s, C, W)$ 
        }
        \Else{
            $L_p', L_n' \gets \textsc{PickDisagreed}(U_s, C, W)$ \label{line:active_learning_end}
        }
        $L_p \gets L_p \cup L_p'$, $L_n \gets L_n \cup L_n'$, $U \gets U - (L_p' \cup L_n')$  \label{line:update_data} \\
        $W \gets \textsc{ComputeScore}(L_p, L_n, C)$ \label{line:W2}
    }
    \caption{UDF selection using active learning.}\label{algorithm}
\end{algorithm}

\subsubsection{UDF selection and active learning}
To best align with the user's intent,
\system strives to select the candidate UDF that yields the best F1 score for a given set of user labels.
As summarized in Algorithm~\ref{algorithm},
\system uses active learning~\cite{DBLP:conf/iccv/MullapudiPMRF21, DBLP:journals/pvldb/ZhangD0HKB23, DBLP:conf/aistats/KarimiGKR0021}
to reduce the number of labeled examples needed from the user. 
This algorithm's hyperparameters include: a labeling budget $b$, thresholds $t_p$ and $t_n$ giving the number of positive and negative samples needed to initiate active learning, and the number of tuples $n_s$ sampled in each iteration. \Cref{sec:evaluation} details the values used in our experiments.

During each iteration, the algorithm randomly samples $n_s$ tuples from the database
(Line~\ref{line:sample}). It operates in two phases: \textit{bootstrapping} and \textit{active learning}. 
The bootstrapping phase
collects at least $t_p$ positive and $t_n$ negative samples. While the number of labeled positives ($|L_p|$) or negatives ($|L_n|$) remains below its respective threshold,
\system selects samples most likely to be positive or negative (Lines \ref{line:bootstrap_start} to \ref{line:bootstrap_end}). Once enough initial labeled data has been collected, \system transitions to active learning, adopting the margin and positive strategy from~\cite{DBLP:conf/iccv/MullapudiPMRF21}. Specifically, if $|L_p| < |L_n|$, it asks the user to label samples most likely to be positive; otherwise, it picks the sample with the greatest disagreement among UDF candidates to differentiate them~\cite{DBLP:journals/pvldb/ZhangD0HKB23} (Lines \ref{line:active_learning_start} to \ref{line:active_learning_end}). 

In $\textsc{PickPositive}(U_s, C, W)$ and $\textsc{PickNegative}(U_s, C, W)$, the algorithm priorities VLM annotations obtained in~\Cref{subsec:data_labeling} over the UDF candidates' majority vote, as VLM labels are empirically more reliable. In $\textsc{PickDisagreed}(U_s, C, W)$, it computes a disagreement score for each sample in $U_s$ based on UDF candidates $C$, selecting the one with the highest disagreement~\cite{DBLP:conf/aistats/KarimiGKR0021, DBLP:journals/pvldb/ZhangD0HKB23}. Each sample's disagreement score is the weighted disagreement among UDF candidates, with candidate weights initialized to $1/|C|$ (Line \ref{line:W1}) and updated to its performance (F1 score in our prototype) over $L_p$ and $L_n$ after each iteration (Line \ref{line:W2}). At the end of each iteration, $L_p$, $L_n$ and $U$ are updated with the new user labels (Line \ref{line:update_data}).

\subsubsection{Dummy UDFs}~\label{subsec:dummy_udf}
Several factors can contribute to the unsatisfactory performance of generated UDFs, including the complexity of the target concept, the low quality of VLM labels, and an insufficient number of positive training samples for rare events. When no UDF candidate performs well, \system should not apply a UDF that can hurt the performance.  
To prevent this, \system appends a \textit{dummy UDF} to its list of UDF candidates. A dummy UDF is a constant-valued function that produces \texttt{True} and is equivalent to omitting the predicate. When all other candidates perform poorly, \system opts for the dummy UDF during its UDF selection process.

\subsubsection{UDF generation strategy}~\label{sec:generation_strategy}
Program-based UDFs and distilled-model UDFs offer different trade-offs in terms of performance, interpretability, generation cost, and inference throughput.
\system provides the following four strategies based on the user's preference and the system's performance requirements: \texttt{program} only generates program-based UDFs, \texttt{model} only generates distilled-model UDFs, \texttt{llm} asks the LLM to decide whether to generate program-based or distilled-model UDFs, and \texttt{both} generates both classes of UDFs. 
\iftechreport
\Cref{fig:deciding_udf_type_prompt} shows an example prompt of using the \texttt{llm} UDF generation strategy.
\fi 
Depending on the user specification, one or more program-based UDFs may be generated for each proposed UDF, and one distilled-model UDF is generated for each proposed UDF. By default, \system uses the \texttt{both} strategy to maximize query performance. In~\Cref{subsec:exp_udf_selection}, we also evaluate the performance of using the \texttt{llm} strategy to automatically choose UDF types. 

\subsection{\revision{Comparing with direct LLM methods}}~\label{subsec:llm}

\revision{
As an alternative to \system, a multimodal LLM can directly answer user queries. However, as shown in prior work~\cite{DBLP:conf/cvpr/MaHGGGK23, DBLP:journals/corr/abs-2406-14852} and as we confirm in~\Cref{subsec:exp_llm_method}, LLMs still struggle with \textit{compositional} queries, and directly providing the LLM with the video frames and the user query in a single prompt (the ``\textbf{vanilla LLM}'' approach in ~\Cref{subsec:exp_llm_method}) leads to low F1 scores. Furthermore, the cost scales with the dataset size, making it expensive for large video collections today, as we discussed in~\Cref{sec:distilled_model_udfs} and
evaluate in~\Cref{table:cost}.
}

\revision{
A higher performance LLM baseline is to parse the query into our DSL, then use a combination of existing cheap models for common concepts and an LLM for application-specific concepts (i.e., \system without UDF generation, or ``\textbf{LLM for concepts}'' in ~\Cref{table:cost}). 
While it may yield good F1 scores, this approach is only practical for short videos and a small number of queries. 
For larger workloads, the cost and latency of this method are even greater than ``vanilla LLM'' due to repeated LLM calls to identify concepts for each object or object pair. As~\Cref{table:cost} shows, executing just one query with three missing UDFs over the CLEVRER dataset using GPT-4o costs about \$1,504 USD.
Reducing costs through sampling is also non-trivial because low sampling rates can degrade query performance, especially for compositional events involving temporal dimensions.
As we show in~\Cref{subfig:sampling_rate}, even a small downsampling factor of 8 can greatly reduce the upper bound of achievable F1 scores from 1.0 to 0.51. 
}

\revision{
\system overcomes these issues by using the multimodal LLM to learn each new application-specific concept once, over a fixed-size sample of the data, and generate cheaper, domain-specific models (either program-based UDFs or distilled-model UDFs) that can then be applied to the remainder of the videos and to all future queries, avoiding repeated reliance on the LLM and reducing overall costs. 
}
\section{Implementation}~\label{sec:implementation}

\system is implemented in Python using the AutoGen framework~\cite{wu2023autogen}. The Query Executor converts queries in DSL notation into SQL and uses a relational engine (DuckDB~\cite{Raasveldt_DuckDB} in our prototype) to execute them. We apply the same query translation algorithm as in~\cite{DBLP:journals/pvldb/ZhangD0HKB23} to optimize query execution. To reduce system latency, we parallelize  API calls to the LLM.

\system requires a preprocessing stage per dataset before queries can be issued.
The user provides a video dataset and optionally a set of UDFs. If these UDFs include object detection and tracking models, \system uses them to populate the \texttt{Objects} relation. Otherwise, \system uses a predefined object detection and tracking model to identify common objects. 
\system pre-extracts features into Parquet files for use by distilled models at query time (\Cref{subsec:model_training}), and we utilize NVIDIA DALI~\cite{nvidiadali} to accelerate this process.
\system also executes the initial UDFs ahead of time to populate the \texttt{Relationships} and \texttt{Attributes} relations and compute the active domains for \texttt{oname}, \texttt{rname}, and \texttt{aname}. These operations happen as soon as \system receives the videos and UDFs and before the user issues any queries. If no initial UDFs are supplied, the \texttt{Relationships} and \texttt{Attributes} relations will be empty.  
Similarly, when a new UDF is later created, our prototype runs it over the entire dataset and materializes the results in the database. It then converts the UDF into a value-lookup UDF so that subsequent queries can reuse the materialized results.
All UDFs are implemented in Python.

There is preprocessing overhead in the initial setup. The cost of running object detection, object tracking, and feature extraction is relatively modest compared to that of LLMs and VLMs, and they can be run locally. 
The cost of executing a UDF depends on its implementation. In this paper, we materialize all results in the video dataset before issuing new queries. 
Several works have focused on optimizing the execution of queries with UDFs~\cite{DBLP:journals/corr/abs-2403-14902, DBLP:conf/sigmod/XuKAR22}, which could be integrated into our system.
\section{Evaluation}~\label{sec:evaluation}

\noindent\textbf{Baselines.} 
We compare \system against VisProg~\cite{DBLP:conf/cvpr/GuptaK23} and EQUI-VOCAL~\cite{DBLP:journals/pvldb/ZhangD0HKB23}, both of which are limited to predefined UDFs when answering queries. VisProg uses a different set of modules for different tasks, so we manually write new modules for the system to be able to answer compositional queries.  These include both logical (e.g., \texttt{Eval}, \texttt{Event}, \texttt{Before}) and conceptual  (e.g., \texttt{Red}, \texttt{Holding}) modules. VisProg also requires in-context examples, which we create separately from the evaluation queries.
While EQUI-VOCAL does not need such examples, it does not accept NL queries as input.  To address this, we adapt \system's approach of converting NL into the EQUI-VOCAL DSL using an LLM.
\revision{We further compare \system with two LLM baselines as described in~\Cref{subsec:llm}.}

\noindent\textbf{Metrics.} We evaluate query answering performance using \revision{F1 scores, precision, and recall}. We propose and generate UDFs using training data and report query \revision{performance} over the test set. We evaluate each dataset using 30 queries, and each query is run three times. 
\revision{We additionally evaluate the system efficiency in terms of latency and cost. For latency, we report the wall-clock time as perceived by the user from the moment they submit a query to the moment when \system returns the query results. For cost, we measure both the LLM API invocation cost using the OpenAI pricing model~\cite{openai_pricing} and the resource cost using the AWS pricing model~\cite{aws_spot,aws_bedrock}.}

\noindent\textbf{Datasets:} 
\revision{
We evaluate \system on three datasets: CLEVRER~\cite{CLEVRER2020ICLR}, CityFlow-NL~\cite{DBLP:journals/corr/abs-2101-04741} and Charades~\cite{DBLP:conf/cvpr/JiK0N20}. CLEVRER is a benchmark that facilitates testing queries with varying complexities. CityFlow-NL and Charades represent real-world applications---traffic monitoring and human activities---typical for video data management systems. 
UDFs in CityFlow-NL include fine-grained vehicle types,
while Charades comes with complex human interactions.
The evaluation queries are based on concepts from these datasets, which are labeled by the original annotators and reflect events from those domains.
}

\noindent\textit{CLEVRER:} The CLEVRER~\cite{CLEVRER2020ICLR} dataset consists of 10,000, 5-second synthetic videos of moving objects. To determine ground-truth information, following~\cite{CLEVRER2020ICLR}, we use a Mask R-CNN~\cite{DBLP:conf/iccv/HeGDG17} to locate objects and predict their colors, shapes, and materials. We write rule-based functions to extract spatial relationships and attributes. For our experiments, we use the same Mask R-CNN and rule-based functions as UDFs, for a total of six relationships and 17 attributes. Among them, we select two relationships and nine attributes as the \textit{base UDFs}, which are available to all systems during query evaluation. We use templates to automatically generate 30 target queries with seven predicates,
\revision{up to three variables (i.e., three distinct objects), up to three region graphs, and duration constraints with three possible values. }
We ensure there are at least 5\% positive examples in the dataset for each target query (we ensure the same for all datasets). We then rewrite the DSL queries as NL queries using GPT-4 and manually verify the correctness.

\noindent\textit{CityFlow-NL:} The CityFlow-NL~\cite{DBLP:journals/corr/abs-2101-04741} dataset contains traffic videos captured from multiple cameras and NL descriptions for vehicle tracks. Following~\cite{DBLP:conf/cvpr/LeNLCCH23}, we extract vehicle colors and types from the NL descriptions. Since only sampled vehicle tracks are annotated, we only consider them in our evaluation. We extract six spatial relationships using rule-based predictors and create 1,473 non-overlapping 50-frame video segments from the original dataset.
To create UDFs for attributes, we train binary image classifiers. For relationships, we use the same rule-based functions as UDFs.
Then, we select three relationships and four attributes as \textit{base UDFs}.
We automatically generate 30 queries with up to four predicates,
\revision{up to three variables, up to three region graphs, and duration constraints with three possible values. 
An example query is "a vehicle o0 with type sedan is to the right of another vehicle o1 for at least 15 frames. Then, o0 is to the left of o1 for at least 15 frames", which could be used to find sedans changing their positions with other vehicles.
}

\noindent\textit{Charades:} The Charades~\cite{DBLP:conf/cvpr/JiK0N20} dataset contains 30-second indoor activity videos. We focus exclusively on frames with scene graph annotations provided by Action Genome~\cite{DBLP:conf/cvpr/JiK0N20}, which only provide human-object relationships but do not cover attributes or object-object relationships. Thus, our evaluation queries focus on human interactions with objects. At query time, only human-object pairs are sampled to collect more positive examples. Incomplete annotations is a common issue in real-world datasets~\cite{DBLP:journals/corr/abs-2401-12168} like Charades. To address this, we augment the dataset by automatically generating dense spatial relationship annotations using rule-based predictors to replace the existing ones. For semantic relationships, we train binary image classifiers. In total, we have 20 relationships (five spatial, 15 semantic) and 35 object classes. Among them, we select two spatial and seven semantic relationships as the base UDFs. 
We automatically generate 30 target queries with up to four predicates (one is always \texttt{object(o$_i$,`person')}),
\revision{up to three variables, and up to two region graphs, but without duration constraints.
An example query is "a person is eating something, and then the same person is holding another object while being in front of it", which could retrieve videos including a person taking a bite from an apple, and then picking up a cookbook in front of them to read a recipe.
}

For each dataset, we use half of the videos as training data and the rest as test data. To generate ground truth labels, we run each target DSL query on the dataset.

\noindent\textbf{Evaluation setup.} We conduct all experiments on a compute cluster. For each experiment, we request one node with eight Intel Xeon Gold 6230R CPUs at 2.10GHz, 200GB of RAM, and one NVIDIA A40 GPU. We use the GPT-4o model (\texttt{gpt-4o-2024-08-06}) as the LLM and VLM across all systems. We configure \system as follows. For the CLEVRER dataset, we use a labeling budget of 20 for each UDF during selection. We generate 10 candidate programs, allowing numeric parameters and frame pixels as inputs. During UDF selection, we bind each numeric parameter to its default value as well as five random values. When generating distilled-model UDFs, we ask the VLM to annotate 100 sampled frames per UDF. 
Given their increased complexity,
for the CityFlow-NL and Charades datasets, we set the labeling budget per UDF to 50 and the number of VLM-annotated frames per UDF to 500.
100 frames are annotated for the initial training, and then active learning is performed to select 100 additional frames in each round. Frame pixels are disallowed during program-based UDF generation due to their limited impact on performance. 
To focus our evaluation on relationships and attributes, we assume objects have already been detected and tracked.

\subsection{End-to-end performance}~\label{subsec:exp_end_to_end}

\evalAllFigure

\subsubsection{\revision{F1 score, precision, and recall.}}~\label{subsec:end_f1_score}
We first evaluate the end-to-end performance of \system and baseline systems by varying the number of missing UDFs from zero (i.e., all required UDFs are available) to three. In each experiment, all base UDFs are available, but $X$ supplemental UDFs are randomly removed. We execute thirty distinct queries three times.
\Cref{fig:eval_all} presents the \revision{F1 scores, precision, and recall}. \textbf{\system maintains high \revision{performance} even when UDFs are missing.} When all required UDFs are available, all systems perform similarly.  
For CLEVRER, F1 scores with no missing UDFs reach 1.0 since predefined UDFs are also used to generate ground-truth scene graph annotations.
However, CityFlow-NL and Charades contain less accurate ML models in their UDFs, leading to lower F1 scores. When UDFs are missing, VisProg and EQUI-VOCAL experience significant F1 score drops, as they cannot generate new UDFs, with VisProg performing worse due to its higher likelihood of including unavailable UDFs. \system mitigates F1 score degradation by generating new UDFs as needed, though performance drops more on Charades as its missing UDFs involve complex, harder-to-generate semantic relationships.
\revision{Interestingly, while EQUI-VOCAL achieves higher recall than \system on CLEVRER and CityFlow-NL when UDFs are missing, it does so at the expense of substantially lower precision, for an overall lower F1 score.}

\subsubsection{\revision{Comparing with direct LLM methods}}~\label{subsec:exp_llm_method}

\vanillaLlmFigure

To understand the performance gap between \system and using an LLM directly,
\revision{
we evaluate both approaches on CLEVRER. We use the same queries from the end-to-end experiments, but under a simplified setting due to the high expense of GPT-4o.
}
We select 500 videos and downsample each by 75\%, retaining every fourth frame. We use the first 10 queries as evaluated in~\Cref{subsec:end_f1_score} but remove any duration constraints due to the downsampling. We compare \system to a vanilla LLM method in which we provide GPT-4o with a sequence of video frames and ask \revision{
with the prompt: ``Examine the sequence of frames from a video and determine if the following event occurs? <user\_query>. Answer with `yes' or `no'.''
}
We use OpenAI's Batch API that offers 50\% lower cost. \Cref{subfig:llm_baseline} shows that \textbf{\system achieves significantly higher F1 scores than the vanilla LLM approach.} The cost of the vanilla LLM method scales with the dataset size, averaging \$5.17 USD per query under this simplified setting. If extended to the full 5,000 videos in the test set of CLEVRER, the estimated cost would be \$207 USD per query.

\revision{
As an alternative baseline, one can use an LLM to classify new concepts instead of generating UDFs. However, as discussed later in~\Cref{subsec:exp_cost}, the cost of doing so without downsampling is prohibitively high. While downsampling reduces LLM costs, it can also degrade query performance. 
\Cref{subfig:sampling_rate} shows the F1 scores of executing ground-truth DSL queries (i.e., assuming perfect LLM predictions) over the simplified CLEVRER with various downsampling factors. \textbf{Downsampling greatly reduces the upper bound on achievable F1 scores, with performance dropping to 0.51 at a downsampling factor of just 8.}
}

\EfficiencyFigure

\subsubsection{\revision{Execution time.}}~\label{subsec:exp_time}
To analyze \system's efficiency, we perform an end-to-end evaluation of execution time. \Cref{subfig:total_time} breaks down the wall-clock time for each system component, with each stacked bar showing the mean execution time for queries with different numbers of UDFs that \system \textit{decides to generate}.
\revision{\system may occasionally decide to generate extra, unnecessary UDFs. For clarity, we omit CLEVRER with four generated UDFs and CityFlow-NL with three generated UDFs from~\Cref{fig:system_efficiency}, as each has only one data point---arising from the LLM's inaccurate decision---and does not reflect an average execution time for that number of generated UDFs. 
}
\textbf{The average execution time is under 100 seconds when all required UDFs are available and increases proportionally with the number of generated UDFs.} In CLEVRER, query execution takes longer when new UDFs need to be created, requiring several minutes to process the full 14-hour video dataset as \system materializes new UDF outputs. Subsequent queries using the same UDFs, however, are as fast as the case with no missing UDFs.
CityFlow-NL and Charades are faster due to smaller datasets and the exclusion of frame pixels in program-based UDF generation, which eliminates the need for video loading and reduces the program execution cost.
For CityFlow-NL and Charades, UDF generation dominates the execution time, as the number of VLM-annotated frames per UDF increases to 500, with active learning applied after labeling every 100 samples. UDF selection requires non-negligible time across all datasets, involving tasks such as frame loading, feature extraction, program execution, and model prediction.

\Cref{subfig:udf_generation_time} provides a detailed breakdown of wall-clock time for the UDF generation process. Since \system asynchronously parallelizes calls to the LLM, we measure the time breakdown as follows. All categories except for ``other'' capture the synchronous portions of each sub-step (i.e., excluding LLM calls), while ``other'' represents the remaining observed time, which includes the time spent explicitly waiting on asynchronous LLM calls. Program-based UDF generation is labeled as ``program,'' while distilled-model UDF generation is further divided into sub-steps (``init'', ``data loading'', etc.). \textbf{Distilled-model UDF generation takes significantly longer than program-based UDFs, but no single sub-step consistently dominates across all cases.} Specifically, training distilled models takes the longest in CLEVRER and CityFlow-NL, whereas active learning takes the longest in Charades due to a larger dataset. The active learning step could be accelerated by performing it on a sample of the unlabeled dataset. 
The results also suggest that \textbf{LLM calls do not dominate UDF generation time}, thanks to asynchronous execution. For example, during data labeling, \system submits batches of 100 asynchronous tasks to maximize parallelization.

\subsubsection{\revision{Monetary cost.}}~\label{subsec:exp_cost}
\revision{To understand the query execution cost, we consider both the LLM API invocation cost and the compute resource cost.} \Cref{subfig:cost} shows the mean monetary cost of \revision{OpenAI} LLM calls per query. \textbf{\system costs less than \$1.0 USD per query, with data labeling for distilled-model UDFs being the primary cost factor.} Moreover, GPT-4o can be replaced by open-source models. For instance, Amazon Bedrock offers Llama 3.1 70B for a price of \$0.72 USD per 1M tokens~\cite{aws_bedrock}, which is $3.5\times$ cheaper for input and $14\times$ cheaper for output tokens than GPT-4o. Other open-source models and hosting options~\cite{llama_gpt_cmp} are available to further reduce costs, with a potential trade-off in query F1 scores.
\revision{Compute resource costs depend on deployment scenarios. 
For example, on a \texttt{g4dn.12xlarge} EC2 spot instance, the estimated average resource cost of running a query with the largest number of missing UDFs is \$0.7 USD for CLEVRER, \$0.4 for CityFlow-NL, and \$0.5 for Charades~\cite{aws_spot}. Consequently, \textbf{the total per-query cost for each dataset remains under \$1.5.}
}

\costTable

\revision{
\Cref{table:cost} compares the estimated LLM costs of \system, ``vanilla LLM'', and ``LLM for concepts'' (as described in~\Cref{subsec:llm}) using both GPT-4o and Llama 3.1 70B. 
Due to the high expense of the baselines, we estimated costs rather than running full experiments. 
We estimate the cost of ``LLM for concepts'' as follows: Each CLEVRER video contains an average of 544 objects and 1,933 object pairs. Each LLM call requires 300 input tokens (255 for the image and 45 for the task) and one output token (``yes'' or ``no''). We first run the query with missing UDFs removed, then invoke the LLM only on the matching videos to minimize costs. As shown in \Cref{table:cost}, \textbf{\system is much more cost-effective than direct LLM baselines.} While further optimizations---e.g., evaluating immutable attributes on a single frame and propagating them to other frames for the same object---could reduce the cost of ``LLM for concepts'', the cost remains proportional to the dataset size, in contrast to \system.
}

\subsection{UDF proposal}~\label{subsec:exp_propose_udfs}

\nltodslTable

Translating NL queries to DSL is not a contribution of this paper, but we report it as it affects the end-to-end performance. For this experiment, all UDFs are available, and the LLM translates the same queries from the end-to-end experiment, running each three times. We measure the F1 score of the generated queries. As shown in \Cref{table:nl_to_dsl}, \textbf{\system can effectively parse the complex textual queries into the DSL notation}, returning correct video segments at least 79\% of the time, which increases to ${\ge}97\%$ if we consider an F1 score $\geq$ 0.98. Common translation errors include missing and redundant predicates, misuse of UDFs with similar names, and incorrect temporal ordering of predicates.

\proposingUdfsTable
We evaluate \system 's UDF proposal using the same set-up as the end-to-end evaluation. A proposed UDF is considered correct if its name matches one of the target query's ground truth UDFs, with a manual check for synonyms. As shown in~\Cref{table:proposing_udfs}, \textbf{\system proposes fewer UDFs than expected, with more false negatives (FNs) than false positives (FPs).} This occurs because some new UDFs are equivalent to available UDFs (e.g., ``above(o1, o2)'' vs. ``beneath(o2, o1)''). However, this is beneficial as it avoids unnecessary UDF generation. Most other FNs are mainly because \system decides to approximate using other available UDFs (e.g., ``touching'' for ``holding'').
\textbf{Most FPs arise from proposing compositional UDFs}, such as ``behind\_and\_near'' for ``behind'' and ``near''.

\subsection{UDF generation}~\label{subsec:exp_udf_generation}

\programUdfPerformanceTable

We now evaluate the performance of program-based UDFs. We consider all correctly proposed program-based UDFs across the 270 experiments with the largest number of new UDFs (3 datasets $\times$ 30 queries $\times$ 3 runs). We classify UDFs into two categories, one where at least one of their best implementations is program-based, and another where the best is not program-based. 
\Cref{table:program_udf_performance} shows the median F1 score of the best generated program for each UDF (best UDFs) and all generated programs for each UDF (all UDFs). \textbf{When the best UDF generation is program-based, \system consistently produces high-quality programs.} However, not all generated UDF candidates are of high quality; for instance, the median F1 score of all program-based UDFs for Charades is only $0.092$. Thus, \system must carefully select the best candidate during the UDF selection phase. When the best UDF is not program-based, F1 scores decrease significantly, indicating that certain relationships or attributes may not be well-suited to program-based UDFs. 

\iftechreport
\programUdfTypesTable

As described in~\Cref{subsec:program_based_udfs}, \system supports program generation with various enhanced features. We evaluate how often \system uses these features by categorizing program-based UDFs into four types: ``all'' (represents all generated programs), ``reuse'' (utilizes results from existing UDFs), ``param'' (takes numeric parameters as inputs), and ``pixel'' (accepts frame pixels as inputs). \Cref{table:program_udf_types} shows the number of best generated programs for each UDF in each category. The results indicate that the best programs span various types, \textbf{highlighting the importance of enhanced features in generating high-quality program-based UDFs}. 
\fi

\modelUdfPerformanceTable

Using the same method as above, we evaluate the performance of distilled-model UDFs. UDFs are classified into two categories, one where at least one of their best implementations is distilled-model, and another where none are distilled-model. 
\Cref{table:model_udf_performance} shows the median F1 score for both distilled-model generation and dummy generation (as the baseline) for each UDF. \textbf{When the best UDF generated is distilled model, \system shows a significant improvement in F1 scores over the baseline.} When the best UDF is not distilled model, the F1 scores of distilled models are comparable or worse than the baseline, indicating that certain relationships or attributes may not be suitable for distilled-model UDFs. For CityFlow-NL, F1 scores are higher for UDFs whose best implementations are not distilled models, likely due to the complexity of semantic attributes that are more challenging to classify.

\intentAmbiguityFigure

\revision{
To access \system's ability to handle ambiguous intent in program-based UDFs, we test three concepts from CLEVRER---``behind'', ``far'', and ``location\_bottom''. For each concept, we execute the same query with one predicate but three different interpretations. For instance, ``behind'' is interpreted as: (1) o1's center above o2's center, (2) o1's center below o2's center, and (3) o1's bottom edge above o2's bottom edge. 
We compare \system to a baseline that uses an LLM to generate a single program without numeric hyperparameters or UDF reuse. \Cref{fig:intent_ambiguity} shows the F1 scores for each concept across varying interpretations, showing that \textbf{\system can more robustly adapt to different definitions of the same concept}. 
The baseline struggles with ``behind'' since it uses an interpretation based on bounding box overlap that mismatches all three ground truths. While it works well for one interpretation of the other concepts, its single-generation method fails to accommodate varying interpretations, leading to significant performance drops. 
}
\revision{
One limitation of \system is that a concept becomes fixed once a UDF is generated. The UDF would have to be explicitly deleted or disabled if a user wanted a different interpretation for an already-existing concept. Additionally, \system currently does not handle intent ambiguity for distilled-model UDFs, but this can be extended by asking LLMs to generate diverse descriptions of the target concept and distilling multiple models.
}

\modelUdfLabelingQualityTable

We further examine the VLM's labeling quality during distilled-model UDF generation on the CityFlow-NL and Charades datasets. We evaluate nine attributes in CityFlow-NL and 18 relationships
in Charades. For each attribute or relationship, we use GPT-4o to label a randomly selected, balanced set of 500 samples. \Cref{table:model_udf_labeling_quality} shows the average F1 score of each concept over three runs. \textbf{GPT-4o generally achieves high labeling quality for a wide range of concepts}: seven out of the 27 concepts achieve F1 scores of at least 0.9, and 15 attain F1 scores of at least 0.8. However, GPT-4o struggles to label the five spatial relationships, suggesting that \textbf{VLMs like GPT-4o are still limited in spatial reasoning}.

\subsection{UDF selection}~\label{subsec:exp_udf_selection}

\UdfSelectionTable

\revision{\Cref{table:udf_selection} reports the UDF selection results.} 
The ``best'' column shows the number of UDFs selected by \system that achieve the highest F1 score among all UDF candidates, while the ``80\% of best'' column shows how many attain at least 80\% of the best F1 score. \textbf{\system effectively selects better-performing UDFs from candidates, even with a labeling budget as low as 20.} Although \system does not always pick the best UDF due to similar scores for many candidate UDFs, it can still select a good UDF implementation at least 85\% of the time (with an F1 score of at least 80\% of the best implementation).
\revision{
\Cref{table:udf_selection} also breaks down the types of UDFs that \system selects. \textbf{To handle diverse semantic concepts, \system generates and selects UDFs of various types}, underscoring the necessity for \system to support two different types of UDFs. Interestingly, \system selects more dummy UDFs (41 instances, 28\%) on the Charades dataset than the other datasets. Notably, 20 instances are ``holding'', which is difficult to distinguish under the current labeling budget. Another 17 cases are ``in'', which our rule-based predictor identifies by checking bounding box overlap. This method classifies most object pairs as having this relationship, thereby allowing even the dummy UDFs to perform exceptionally well. 
}

\choosingUdfTypeTable

\system supports four UDF generation strategies, as described in~\Cref{sec:generation_strategy}. We now compare the ``\texttt{both}'' and ``\texttt{llm}'' strategies in selecting the correct UDF type.
When multiple candidates of different types share the highest F1 score, any of these types is considered correct. Using the same end-to-end experiment setting with the largest number of missing UDFs,
\Cref{table:choosing_udf_type} shows the number of correctly selected UDF types. The results are divided into two categories: one where the best UDF type is not ``dummy'' and one where it is ``dummy.''
When the best UDF type is not ``dummy,'' \textbf{\system correctly selects the UDF type at least 72\% of the time with the ``\texttt{both}'' strategy}, reaching up to 90\% on CLEVRER. 
Interestingly, \textbf{using ``\texttt{llm}'' with GPT-4o yields a higher accuracy than ``\texttt{both}''}. However, this strategy is highly sensitive to the model used; switching to an earlier GPT-4 Turbo model reduces the accuracy to 46--64\%. Despite this, the ``\texttt{llm}'' strategy demonstrates potential to further improve performance while reducing latency and cost.
When the best UDF type is ``dummy,'' \system can also select the correct UDF type 43\% to 90\% of the time using the ``\texttt{both}'' strategy.

\udfSelectionFigure

We study the impact of active learning and dummy UDFs to the UDF selection process, using the same end-to-end experiment setting with the largest number of new UDFs. 
\Cref{fig:udf_selection} compares F1 scores across three system variants: the complete system with both active learning and dummy UDFs (All), the system with random sampling and dummy UDFs (No active learning), and the system with active learning but without dummy UDFs (No dummy UDFs). \textbf{Both active learning and dummy UDFs help achieve higher F1 scores.} \revision{
For CityFlow-NL, the best UDF implementations are easily distinguishable from candidates, so active learning does not improve F1 scores but also does not degrade system performance.
}

\initialLabelingFigure

\revision{
\system reduces user effort by using active learning to request labels at query time rather than asking the user to provide all examples upfront. \Cref{fig:init_label} compares the number of samples needed under two approaches---active learning during UDF selection and random sampling before query execution---to obtain at least ten positive examples for a given UDF. We consider all generated UDFs from the end-to-end experiment and classify them as ``easy'' (spatial relationships and attributes with relatively balanced labels) or ``hard'' (other concepts with greater class imbalance). \textbf{Active learning allows \system to collect positive samples with reduced labeling effort, particularly for ``hard'' concepts.} 
Since most object pairs exhibit ``in'' relationships in Charades, random sampling requires labeling only 11.5 samples on average to yield 10 positives. As a result, the overall labeling effort for ``easy'' concepts in Charades is similar between random sampling and active learning.
}

\section{Related work}

\textbf{Video analytics.} Numerous video analytics systems have been developed to support a wide range of data management tasks~\cite{DBLP:journals/pvldb/DaumZ0MHKB23, DBLP:conf/cidr/DaumZHBHKCW22, DBLP:journals/corr/abs-2305-03785, DBLP:journals/corr/abs-2308-03276, DBLP:journals/pvldb/KossmannWLTCKM23}. Query execution over videos typically involves running expensive ML models. Thus, many techniques have been proposed to accelerate query processing, including indexing~\cite{DBLP:conf/sigmod/KangGBHZ22, DBLP:conf/sigmod/HeASC20, DBLP:conf/sigmod/HuGH22}, sampling~\cite{Moll2020ExSampleES, DBLP:conf/sigmod/BastaniHBGABCKM20, DBLP:journals/pvldb/BangKCMA23}, pre-filtering frames~\cite{DBLP:conf/sigmod/LuCKC18, DBLP:journals/pvldb/HeDCB21, he2023masksearch, DBLP:conf/sigmod/XarchakosK19}, reusing results~\cite{DBLP:conf/sigmod/XuKAR22}, and building specialized models~\cite{DBLP:journals/pvldb/KangBZ19, DBLP:conf/icde/AndersonCRW19}. \system can incorporate existing methods to optimize query execution. 

\textbf{Compositional video query processing.} \system is most related to systems designed for compositional video queries~\cite{DBLP:journals/pvldb/BastaniMM20, DBLP:journals/pvldb/ChenKYY22, DBLP:journals/corr/abs-1910-02993, DBLP:conf/sensys/LiuGUMCG19, DBLP:journals/corr/abs-2104-06142, DBLP:conf/bigdataconf/YadavC19, DBLP:conf/cav/MellBZB23, DBLP:journals/pvldb/ZhangD0HKB23, DBLP:conf/mlsys/YuZCXZWPL024, DBLP:conf/sigmod/XuKAR22}. However, these systems often require users to have a certain level of database expertise to manually construct compositional queries~\cite{DBLP:journals/pvldb/ChenKYY22, DBLP:journals/corr/abs-1910-02993, DBLP:conf/sensys/LiuGUMCG19, DBLP:conf/sigmod/XuKAR22, DBLP:conf/mlsys/YuZCXZWPL024} or to provide examples to learn a query from~\cite{DBLP:journals/pvldb/ZhangD0HKB23, DBLP:conf/cav/MellBZB23}. In contrast, \system leverages advances in LLMs and allows expressing queries in NL. 

\textbf{LLMs with tools.} LLMs are widely used to tackle challenging text tasks across a variety of applications~\cite{chen2023seed, DBLP:conf/nips/GuoGNLGCW023, DBLP:conf/www/0004LMAA21, DBLP:conf/chi/PangSJR24, DBLP:journals/pvldb/KayaliLFVOS24, DBLP:journals/corr/abs-2107-03374, wang2025jupybaraoperationalizingdesignspace, wang2025dracogpt}. By integrating external tools, LLMs can address even more complex reasoning tasks~\cite{DBLP:conf/nips/LuPCGCWZG23, DBLP:conf/nips/SchickDDRLHZCS23, DBLP:conf/iclr/QinLYZYLLCTQZHT24, DBLP:journals/corr/abs-2307-13854, ma2024mms}, including vision tasks~\cite{MitraCCoT, DBLP:journals/corr/abs-2312-00937, DBLP:conf/cvpr/GuptaK23, DBLP:conf/iccv/SurisMV23, DBLP:conf/nips/0001ST00Z23}. However, these systems generally rely on the availability of existing tools or modules. For instance, VisProg~\cite{DBLP:conf/cvpr/GuptaK23} and ProViQ~\cite{DBLP:journals/corr/abs-2312-00937} utilize LLMs to transform complex tasks into executable programs that invoke predefined tools. \system also employs LLMs to handle compositional queries over videos but extends this capability by generating new UDFs. Several systems explore the capacity of LLMs to create new tools. LATM~\cite{DBLP:journals/corr/abs-2305-17126} generates reusable code snippets for NL tasks, while GENOME~\cite{Chen2023GENOMEGN} generates and reuses code-based modules to solve visual tasks. \system distinguishes itself as an end-to-end VDBMS that generates both program-based and distilled-model UDFs which significantly enhances performance.

\section{Limitations}~\label{sec:limitation}

\system does not currently support relationship and attribute UDFs involving state changes over time, such as identifying a car ``moving close'' to a truck, which requires distance comparisons at different timestamps. As an approximation, multiple frame-level UDFs can represent each state separately; for instance, detecting a ``far'' relationship followed by a ``near'' one.

Materializing and reusing UDF results can be problematic when the definition of an attribute or relationship is not objective (e.g., ``near'') and varies from query to query. To address this, \system could allow users to access and manage generated UDF descriptions, implementations, and results, enabling them to dynamically enable or disable specific UDFs before issuing queries. 

\revision{
\system's performance is affected by its ability to semantically understand the user queries and available UDFs. We assume that predefined UDFs are relevant to the target domain and that an LLM is capable of disambiguating each UDF effectively. The performance of \system thus depends significantly on the underlying LLM, which is why we used the latest LLMs (e.g., GPT-4o) to ensure high-quality semantic understanding. Furthermore, \system currently supports queries with detailed and explicit descriptions. Extending support to vague queries is left for future work.
}

\system leverages LLMs in various system components, making the quality of query results dependent on the underlying LLM performance. To enhance the reliability of LLM-generated answers, \system employs syntax verification, generates multiple candidate UDFs, and leverages user annotations via active learning to select the best implementation. Introducing dummy UDFs further ensures that newly generated UDFs do not negatively impact query results. While users already provide annotations to guide UDF selection, \system could be extended to incorporate additional human interventions, allowing users to access intermediate results for manual examination~\cite{DBLP:conf/uist/YangBMD23, zhang23equivocaldemo, cho24dsg}. Our experiments empirically demonstrate the promising potential of our LLM-based approach, and we posit that as LLMs advance, \system performance will correspondingly improve.
\section{Conclusion}

This paper presents \system, a new system that supports compositional video queries with the capability to generate new UDFs. \system utilizes LLMs to parse natural language queries and automatically determine the need for new UDFs. It supports both program-based and distilled-model UDF generations and improves UDF quality through syntax verification and semantic verification.

\end{sloppypar}
\newcommand{\parsingQueryPrompt}{
    \centering
    \begin{tcolorbox}[enhanced,breakable,attach boxed title to top center={yshift=-3mm,yshifttext=-1mm},
        colback=white,colframe=black,colbacktitle=gray,
        title=Prompt of query parsing,
        boxed title style={size=small,colframe=black}]
        \textbf{Legend}: \highlight{hlc4}{DSL definition}, \highlight{hlc2}{UDF definition}, \highlight{hlc3}{Registered UDF}, \highlight{hlc1}{Instruction}\\

        \highlight{hlc4}{Each video segment is a sequence of N frames. The visual content of each frame is represented by a region graph: A region graph contains a set of objects in a frame, along with a set of relationships between those objects. Objects can optionally have attributes. In our DSL, we use a variable o to represent an object in a query. Different variables represent different objects. All predicates of a region graph are connected by commas. Then, region graphs are connected in temporal sequence with semicolons. Region graphs that appear earlier in the sequence represent temporally earlier frames in the video. We further use the notation Duration(g, d) to require that the region graph g exist in at least d consecutive frames. Negation operation is not supported in our DSL. Remember to always add parentheses around comma-connected predicates. Assume the video segments capture 25 frames per second.} 

        \highlight{hlc2}{A function can take one of the following three formats, depending on if it is a relationship predicate or an attribute predicate: \\
        - relationship predicate: relationshipName(o0, o1). For example, jumping\_in(o0, o1) checks whether o0 is jumping in o1. \\
        - attribute predicate: key\_value(o0). For example, color\_bronze(o0) checks whether the color of o0 is bronze.}

        \highlight{hlc3}{You have access to the following functions: \\
        left\_of(o0, o1): Whether o0 is on the left of o1. \\
        front\_of(o0, o1): Whether o0 is in front of o1. \\
        location\_left(o0): Whether o0 is on the left of the frame. \\
        location\_top(o0): Whether o0 is at the top of the frame. \\
        color\_gray(o0): Whether the color of o0 is gray. \\
        color\_red(o0): Whether the color of o0 is red. \\
        color\_blue(o0): Whether the color of o0 is blue. \\
        color\_green(o0): Whether the color of o0 is green. \\
        shape\_cube(o0): Whether the shape of o0 is cube. \\
        shape\_sphere(o0): Whether the shape of o0 is sphere. \\
        material\_rubber(o0): Whether the material of o0 is rubber.
        }

        \highlight{hlc1}{For text-to-DSL translation tasks, only use the functions you have been provided with. Reply PARSE\_YES when the text is successfully translated into the DSL and verified by the provided function, or PARSE\_NO if parsing the user input requires new predicates that are not listed in the current functions list. The predicates MUST be selected from the provided functions.}
    \end{tcolorbox}
    \noindent\begin{minipage}{\columnwidth}
    \captionof{figure}{Prompt of query parsing.}~\label{fig:parsing_query_prompt}
    \end{minipage}
}

\newcommand{\proposingUDFPrompt}{
    \centering
    \begin{tcolorbox}[enhanced,breakable,attach boxed title to top center={yshift=-3mm,yshifttext=-1mm},
        colback=white,colframe=black,colbacktitle=gray,
        title=Prompt of UDF proposal,
        boxed title style={size=small,colframe=black}]
        \textbf{Legend}: \highlight{hlc1}{Instruction}\\

        \highlight{hlc1}{For function proposal tasks, only use the functions you have been provided with. Reply TERMINATE when the task is done. Please propose the new functions that are necessary to parse the user query, and also include a brief description for each proposed function that explains its purpose as described in the query. The function description should always start with the word ``Whether'' and not contain other comments, explanations, or reasoning.\\
        Let's think step by step. Based on the existing functions, determine what new functions are needed. The proposed function must follow the format. Don't propose functions that contain changes in states. If you have those, propose a separate function for each state instead. For example, do not propose a function merge\_from\_A\_into\_B(o0) that checks whether an object o0 merges from lane A to lane B because it contains two states: o0 is in lane A and o0 is in lane B. Instead, replace it with two separate functions: in\_lane\_A(o0) and in\_lane\_B(o0). Propose as few functions as possible while ensuring that the user's intent can be precisely captured.}
    \end{tcolorbox}
    \noindent\begin{minipage}{\columnwidth}
    \captionof{figure}{Prompt of UDF proposal uses the same prompt as in~\Cref{fig:parsing_query_prompt}, but with an updated instruction.}~\label{fig:proposing_udf_prompt}
    \end{minipage}
}

\newcommand{\implementingUDFPrompt}{
    \centering
    \begin{tcolorbox}[enhanced,breakable,attach boxed title to top center={yshift=-3mm,yshifttext=-1mm},
        colback=white,colframe=black,colbacktitle=gray,
        title=Prompt of program-based UDF generation,
        boxed title style={size=small,colframe=black}]
        \textbf{Legend}: \highlight{hlc1}{Instruction}, \highlight{hlc2}{Schema info}, \highlight{hlc3}{Output format}\\

        \highlight{hlc1}{Generate 10 Python functions with different, diverse semantic interpretations for the following Python task. Each generation should include the semantic interpretation and the Python function implementation, formatted as a dictionary. The response should strictly adhere to the formats described below:\\
        - Task: Write a python function called `py\_near(img, o0\_oname, o0\_x1, o0\_y1, o0\_x2, o0\_y2, o0\_anames, o1\_oname, o1\_x1, o1\_y1, o1\_x2, o1\_y2, o1\_anames, o0\_o1\_rnames, o1\_o0\_rnames, height, width, **kwargs)' that determines whether o0 is near o1.\\
        - Each interpretation should offer a different but reasonable understanding of the task, not just superficial differences like variable names. Seek interpretations that vary in logic and conceptual understanding of the task. Consider geometric, visual, and spatial perspectives. Include assumptions or constraints where relevant.\\
        - Prioritize generating functions that are likely to see frequent use, starting with the most common.}

        \highlight{hlc2}{
        - The input to the function contains the following parameters:\\
            - img: np.ndarray of shape (H, W, C). The image is in the RGB color space, where H is the height, W is the width, and C is the number of channels.\\
            - o0\_oname: str. The class name of object o0.\\
            - o0\_x1: int. The x-coordinate of the top-left corner of the bounding box of object o0.\\
            - o0\_y1: int. The y-coordinate of the top-left corner of the bounding box of object o0.\\
            - o0\_x2: int. The x-coordinate of the bottom-right corner of the bounding box of object o0.\\
            - o0\_y2: int. The y-coordinate of the bottom-right corner of the bounding box of object o0.\\
            - o0\_anames: List[str]. The list of attribute names of object o0.\\
            - o1\_oname: str. The class name of object o1.\\
            - o1\_x1: int. The x-coordinate of the top-left corner of the bounding box of object o1.\\
            - o1\_y1: int. The y-coordinate of the top-left corner of the bounding box of object o1.\\
            - o1\_x2: int. The x-coordinate of the bottom-right corner of the bounding box of object o1.\\
            - o1\_y2: int. The y-coordinate of the bottom-right corner of the bounding box of object o1.\\
            - o1\_anames: List[str]. The list of attribute names of object o1.\\
            - o0\_o1\_rnames: List[str]. The list of relationship names between object o0 and object o1, where object o0 is the subject and object o1 is the target.\\
            - o1\_o0\_rnames: List[str]. The list of relationship names between object o1 and object o0, where object o1 is the subject and object o0 is the target.\\
            - height: int. The height of the frame.\\
            - width: int. The width of the frame.\\
            - **kwargs: Optional numeric parameters that can be adjusted as needed.\\
        - Available object names: [`object']\\
        - Available attribute names: [`location\_left', `location\_top', `color\_gray', `color\_red', `color\_blue', `color\_green', `shape\_cube', `shape\_sphere', `material\_rubber']\\
        - Available relationship names: [`left\_of', `front\_of']\\
        - The origin (x, y) = (0, 0) is located at the top left corner. The x axis is oriented from left to right; the y axis is oriented from top to bottom.
        }

        \highlight{hlc1}{- The function should return a boolean value, indicating whether the relationship between the two objects is true or false.}

        \highlight{hlc1}{- Include `**kwargs' in the function's arguments only if necessary. Only arguments of numeric data types are allowed in `**kwargs'. String, boolean, or object data types are not allowed in `**kwargs'.\\
        - You can use any python packages you want (except for sklearn). IT IS LIFE THREATENING THAT you do not use sklearn library. You do not need to install but only import them before using. You can not use supervised-learning method as there is no training data. Though, you can use frozen models if you want.\\
         - The function should only contain the implementation itself, with no other comments, inline comments, syntax highlighter, explanations, reasoning, or dialogue.}

        \highlight{hlc3}{
        - Use the following output format: \\
        \textasciigrave\textasciigrave\textasciigrave json \\
        \{\\
        \hspace*{1em}"answer": [ \\
        \hspace*{1em}\{ \\
        \hspace*{2em}"semantic\_interpretation": "interpretation", \\
        \hspace*{2em}"function\_implementation": "def py\_near(img, o0\_oname, o0\_x1, o0\_y1, o0\_x2, o0\_y2, o0\_anames, o1\_oname, o1\_x1, o1\_y1, o1\_x2, o1\_y2, o1\_anames, o0\_o1\_rnames, o1\_o0\_rnames, height, width, **kwargs):\textbackslash n \quad   \# Your code here", \\
        \hspace*{2em}"kwargs": \{ \\
        \hspace*{3em}"arg\_name1": \{"min": minimum\_value, "max": maximum\_value, "default": default\_value\}, \\
        \hspace*{3em}// Add more arguments as needed. \\
        \hspace*{2em}\}, \\
        \hspace*{1em}\}, \\
        \hspace*{1em}// Add more functions as needed. \\
        ]\} \\
        \textasciigrave\textasciigrave\textasciigrave
        }
    \end{tcolorbox}
    \noindent\begin{minipage}{\columnwidth}
    \captionof{figure}{Prompt of program-based UDF generation.}~\label{fig:implementing_udf_prompt}
    \end{minipage}
}

\newcommand{\decideUDFTypePrompt}{
    \centering
    \begin{tcolorbox}[enhanced,breakable,attach boxed title to top center={yshift=-3mm,yshifttext=-1mm},
        colback=white,colframe=black,colbacktitle=gray,
        title=Prompt of deciding the UDF type,
        boxed title style={size=small,colframe=black}]
        \textbf{Legend}: \highlight{hlc1}{Instruction}, \highlight{hlc2}{Schema info}\\
        
        \highlight{hlc1}{
        You are tasked with creating a solution to determine ``Whether o0 is behind o1''. You can choose to use either a python function or a computer vision model.\\
        1. Python function: This approach is suitable for tasks that can be determined based on any of the following:\\
        - Existing concepts of objects. You can only leverage concepts from the following predefined list:} 
        \highlight{hlc2}{[`object', `left\_of', `front\_of', `location\_left', `location\_top', `color\_gray', `color\_red', `color\_blue', `color\_green', `shape\_cube', `shape\_sphere', `material\_rubber'].}
        \highlight{hlc1}{These concepts are pre-extracted for each object in the image. Concepts not listed are not available.\\
        - Bounding box coordinates of objects.\\
        - Statistical analysis of pixel values in the image using computer vision libraries.\\
        2. Computer vision model: This approach is suitable for tasks that require understanding the visual content and contextual interpretation of the image.\\
        Please specify your choice by responding with 'programUDF' to use the Python function or 'modelUDF' to use the computer vision model. Choose the approach that you believe will achieve the highest accuracy for the task. Consider only the effectiveness of each approach without concern for computational resources, time, or other constraints. Please respond with the answer only, and do not output any other responses or any explanations.
        }
    \end{tcolorbox}
    \noindent\begin{minipage}{\columnwidth}
    \captionof{figure}{Prompt of deciding the UDF type, utilized in the \texttt{llm} UDF generation strategy.}~\label{fig:deciding_udf_type_prompt}
    \end{minipage}
}

\newcommand{\filterObjectsPrompt}{
    \centering
    \begin{tcolorbox}[enhanced,breakable,attach boxed title to top center={yshift=-3mm,yshifttext=-1mm},
        colback=white,colframe=black,colbacktitle=gray,
        title=Prompt of object-aware sampling,
        boxed title style={size=small,colframe=black}]
        \textbf{Legend}: \highlight{hlc1}{Instruction}, \highlight{hlc2}{Object classes}, \highlight{hlc3}{Output format}\\

        \highlight{hlc1}{Given a list of object classes: }
        \highlight{hlc2}{[`person', `bag', `bed', `blanket', `book', `box', `broom', `chair', `closet/cabinet', `clothes', `cup/glass/bottle', `dish', `door', `doorknob', `doorway', `floor', `food', `groceries', `laptop', `light', `medicine', `mirror', `paper/notebook', `phone/camera', `picture', `pillow', `refrigerator', `sandwich', `shelf', `shoe', `sofa/couch', `table', `television', `towel', `vacuum', `window'],} 
        \highlight{hlc1}{and a function "eating(o0, o1)" that determines "Whether o0 is eating o1", assume that objects are chosen from the object classes listed above. Your task is to identify and list all object classes that can possibly be involved in this concept. It's LIFE THREATENING not to remove object classes that can possibly be involved in this concept. \\}
        \highlight{hlc3}{
        Please format your answer in the JSON format shown below: \\
        \textasciigrave\textasciigrave\textasciigrave json \\
        \{"answer": [ \\
        \hspace*{1em}"object\_class1", \\
        \hspace*{1em}"object\_class2", \\
        \hspace*{1em}// Add more object classes as needed. \\
        ]\} \\
        \textasciigrave\textasciigrave\textasciigrave
        }
    \end{tcolorbox}
    \noindent\begin{minipage}{\columnwidth}
    \captionof{figure}{Prompt of object-aware sampling, utilized in distilled-model UDF generation.}~\label{fig:filtering_objects_prompt}
    \end{minipage}
}

\begin{acks}
This work was funded in part by NSF award 2211133. Nicole Sullivan was supported in part by an NSF graduate fellowship (DGE-2140004).
\end{acks}

\bibliographystyle{ACM-Reference-Format}
\bibliography{others, self}

\iftechreport
\appendix
\clearpage
\section{LLM Prompts}

We provide a set of example prompts we use in \system.

\parsingQueryPrompt

\proposingUDFPrompt

\implementingUDFPrompt

\filterObjectsPrompt

\decideUDFTypePrompt

\fi 

\end{document}